\documentclass[preprint,amssymb,amsmath,amsfonts,nobibnotes,nofootinbib]{revtex4}

\usepackage[utf8]{inputenc}
\usepackage[T1]{fontenc}
\usepackage[colorlinks=true,linkcolor=red]{hyperref}
\usepackage{verbatim}
\usepackage{todonotes}

\usepackage{graphicx}
\usepackage{float}
\usepackage[font=small]{caption}
\usepackage[caption=false]{subfig}

\usepackage{bm}
\usepackage{dsfont}
\usepackage{mathtools}

\begin{document}

\newcommand{\textred}{\textcolor{red}}
\newcommand{\textblue}{\textcolor{blue}}

\newcommand{\R}{\mathbb{R}}  
\newcommand{\C}{\mathbb{C}}  
\newcommand{\diff}{\textup{d}} 
\newcommand{\Tr}{\textup{Tr}} 
\newcommand{\conj}[1]{\overline{#1}} 
\newcommand{\diag}{\textup{diag}} 
\newcommand{\mean}{\mathbb{E}} 
\newcommand{\anglemean}[1]{\left\langle #1 \right\rangle}
\newcommand{\prob}[1]{\mathbb{P}\left\{#1\right\}} 
\renewcommand{\mid}{\,\middle\vert\,}

\newcommand{\binomrv}[2]{\textup{Binom}\left(#1\, ,#2\right)}
\newcommand{\EvalAt}[1]{\Bigg\vert_{#1}} 
\newcommand{\Eye}{\mathds{1}} 
\newcommand\indices[1]{\langle #1 \rangle} 
\newcommand{\Gammap}{\Gamma^\prime}
\newcommand{\ah}{a_h}
\newcommand{\al}{a_l}
\newcommand{\betah}{\beta_h}
\newcommand{\betal}{\beta_l}

\newcommand{\bxi}{\bm{\xi}}
\newcommand{\bzeta}{\bm{\zeta}}

\newcommand{\bRup}{\bm{R}^\uparrow}
\newcommand{\bRdown}{\bm{R}^\downarrow}
\newcommand{\bR}{\bm{R}}
\newcommand{\bQ}{\bm{Q}}
\newcommand{\bQup}{\bm{Q}^\uparrow}
\newcommand{\bQdown}{\bm{Q}^\downarrow}
\newcommand{\bW}{\bm{W}}
\newcommand{\bpi}{\bm{\pi}}
\newcommand{\baf}{\bar{f}}
\newcommand{\bad}{\bar{d}}
\newcommand{\frakt}{\mathfrak{t}}
\newcommand{\lc}{\ell_{c}}
\newcommand{\Deltatrans}{\Delta_\textup{trans}}
\newcommand{\Deltamax}{\Delta_\textup{max}}
\newcommand{\badtrans}{\bad_\textup{trans}}
\newcommand{\badmin}{\bad_\textup{min}}
\newcommand{\badmax}{\bad_\textup{max}}
\newcommand{\CMF}{C_\textup{MF}}

\newcommand{\target}{t}
\newcommand{\targetnode}{\target}
\newcommand{\rootnode}{r}
\newcommand{\mrt}{m_{\rootnode \targetnode}}
\newcommand{\SD}{\textup{SD}}
\newcommand{\SP}{\textup{SP}}

\newenvironment{itquote}{%
    \it\begin{quote}}{\end{quote}}



\title{Information retrieval and structural complexity of legal trees}


\author{Yanik-Pascal F\"{o}rster}
\author{Alessia Annibale}
\author{Luca Gamberi}
\author{Evan Tzanis}
\author{Pierpaolo Vivo}

\affiliation{Quantitative and Digital Law Lab, Department of Mathematics, King's College London,
	Strand, WC2R 2LS, London (United Kingdom)}

\homepage[]{Quantlaw.co.uk}


\date{\today}

\begin{abstract}
We introduce a model for the retrieval of  information hidden in legal texts. These are typically organised in a hierarchical (tree) structure, which a reader interested in a given provision needs to explore down to the ``deepest'' level (articles, clauses,...). We assess the structural complexity of legal trees by computing the mean first-passage time a random reader takes to retrieve information planted in the leaves. The reader is assumed to skim through the content of a legal text based on their interests/keywords, and be drawn towards the sought information based on keywords affinity, i.e. how well the Chapters/Section headers of the hierarchy seem to match the informational content of the leaves. Using randomly generated keyword patterns, we investigate the effect of two main features of the text -- the horizontal and vertical coherence -- on the searching time, and consider ways to validate our results using real legal texts. We obtain numerical and analytical results, the latter based on a mean-field approximation on the level of patterns, which lead to an explicit expression for the complexity of legal trees as a function of the structural parameters of the model. Policy implications of our results are briefly discussed. 
\end{abstract}


\maketitle




\section{Introduction\label{sec:intro}}
What is the maximum number of tenants that a UK landlord may let a property of a given size to without incurring in penalties?
Faced with a legal question of this nature, a layperson would naturally resort to consulting the institutional repository \url{www.legislation.gov.uk}, where a quick keyword search (say, ``overcrowding'') would return -- as most recent reference -- the Housing Act 1985. The requested information will eventually be found in the articles 325 \textit{et sqq.}, which can be located after following the path ``UK Public General Acts'', ``Housing Act 1985'', ``Part X: Overcrowding'', ``Definition of Overcrowding'', ``324 Definition of Overcrowding'',
through several part and section headers of the act.

\begin{figure}[H]
    \centering
    \includegraphics{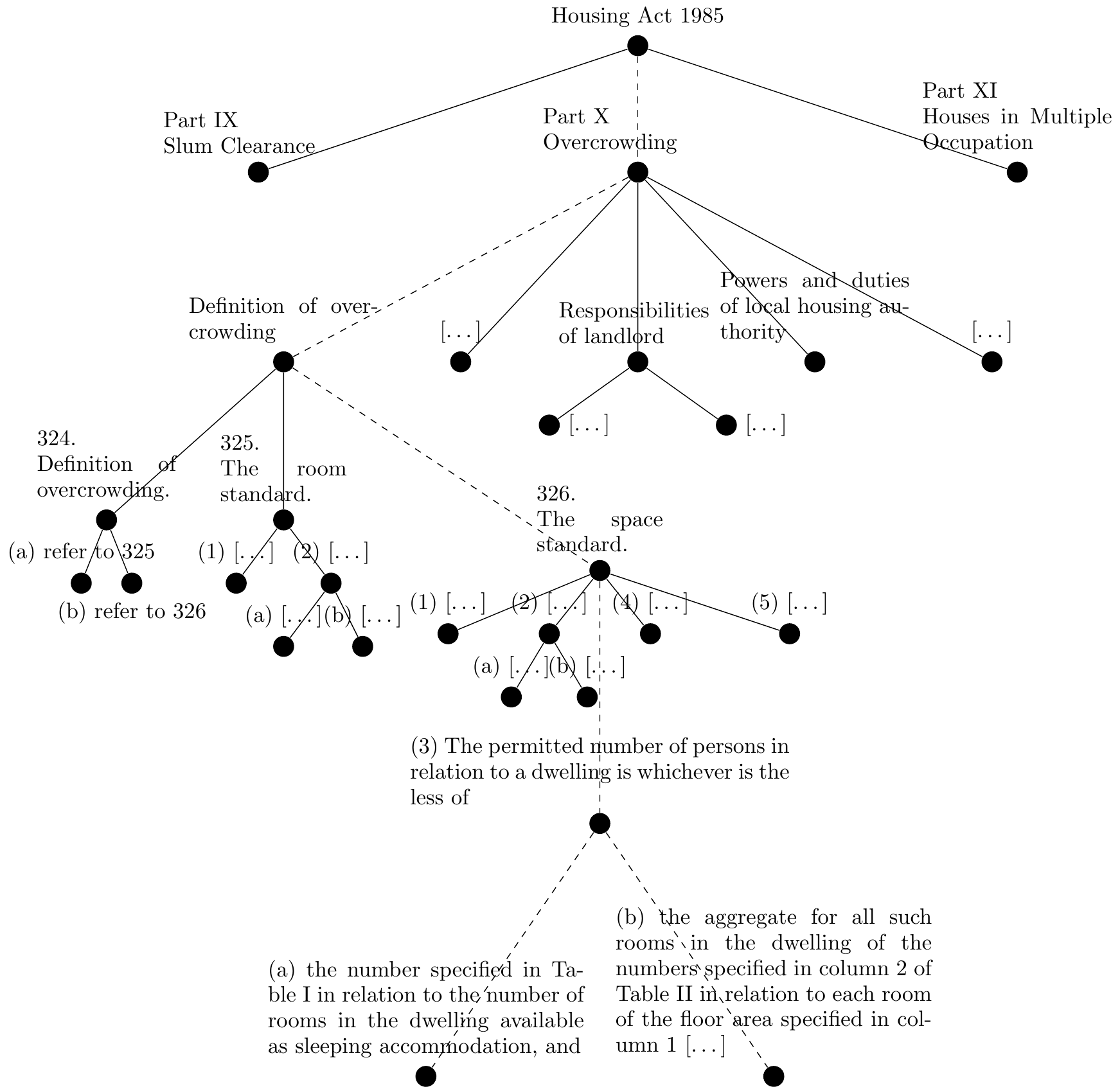} 
    \caption{Example from an excerpt of the UK Housing Act 1985, c.68, to be found at \url{https://www.legislation.gov.uk/ukpga/1985/68/contents}. The nodes of the tree represent structural items of the text such as the Act itself, its Parts, Sections, etc, and two items are linked if one is contained in the other. The dashed edges label the ideal path of a reader researching the question ``What is the maximum number of tenants that a UK landlord may let a property with a given number of rooms to without incurring in penalties?''
    }
    \label{fig:housing_act_example}
\end{figure}

Arguably, a definition of how ``complex'' a piece of legislation is should reflect how fast and reliably information hidden in its text can be retrieved.
The concept of ``legal complexity'' and quantitative measures thereof -- in one of its very many incarnations -- has been considered by legal scholars and -- to a lesser extent -- by scientists in recent years (see next section), so far without reaching a satisfactory and widely accepted consensus on the best framework to use. 
In this paper, we develop a realistic model for the search process of information hidden in a legal text -- organised in a hierarchical fashion -- by a ``typical'' reader who needs to extract a precise answer out of a potentially messy structure of semantic dependencies.   
We consider a tree structure that mimics the organisation of a typical act of Parliament, with primary focus on
the usual structure of legislative bills in the United Kingdom. To standardise labels, we always use the capitalised terms ``Act'', ``Part'', ``Chapter'', ``Section'', ``Article'', and ``Paragraph'' for ease of reference. If applicable, the list can be extended by adding ``sub''-items, e.g. Sub-Sections between Sections and Articles. For the sake of clarity, references to items of this manuscript will be made in small letters and abbreviated (e.g. ``sec.'').
Each item in the hierarchy is identified by a header or set of keywords, which ideally reflect the general content of the corresponding sub-tree: for example, the tree in fig.~\ref{fig:housing_act_example} represents a selection of the Housing Act 1985, with some of the nodes labelled by their textual content. 

Higher up in the hierarchy, the textual content mainly consists of short titles, containing only a small number of keywords (e.g. ``Housing'' and ``Occupation''), while lower nodes often contain full sentences. The Chapter ``Definition of Overcrowding'' of Part X ``Overcrowding'' in fig.~\ref{fig:housing_act_example} contains the text \cite{uk-legislation}
\begin{itquote}
\textbf{324 Definition of overcrowding.}\\
A dwelling is overcrowded for the purposes of this Part when the number of persons sleeping in the dwelling is such as to contravene—\\
(a) the standard specified in section 325 (the room standard), or\\
(b) the standard specified in section 326 (the space standard).
\end{itquote}

Based on the headers/keywords, and 
assuming
they reflect the content underneath, the reader will be more or less inclined to follow a certain path rather than another in their search for a piece of information, planted in one of the leaves of the tree. For the sake of simplicity, we do not consider more complicated network structures with cycles and long-range connections, determined for instance by cross-references or internal amendments. 

We characterise the complexity of a legal tree in terms of the time
a random reader takes on average to reach the sought information by hopping through the nodes of the tree,
guided by the headers' keywords.
The hopping probabilities will reflect the search strategy of the reader and will be defined in sec.~\ref{sec:model}.
The observable we will therefore focus on and give analytical estimates of is the mean first-passage time (MFPT) 
to reach the target starting from the root. 
Leveraging on the MFPTs of random readers of ``legal trees'', we will be able to formulate a closed-form expression for their \emph{structural complexity} in terms of the network parameters, and draw some real-life policy implications for the drafting of legal texts. 

To put our work into context, we review the related literature on complexity of legal systems in section~\ref{sec:literature}.
In sec.~\ref{sec:model} we provide the main definitions of our model for a ``random reader'' hopping through sections of a legal tree according to suitably defined probabilistic rules. Sec. \ref{sec:expected_pattern_distances} covers some analysis of the  statistical properties of this model, with relevant aspects of the theory of MFPTs reviewed in sec.~\ref{sec:mfpt_rev}. Our main results are shown in sec.~\ref{sec:compfunc}: these are based on a mean-field -- or rather, ``mean-text'' -- approximation for the MFPT between the starting point and a target node of our random reader model. This approximation serves as the definition of our ``complexity function'' for legal trees.
We compare analytical and numerical results 
in sec.~\ref{sec:simulations}, and summarise our findings in sec.~\ref{sec:conclusio}. 
Technical derivations of our results are shown in the appendices~\ref{app:expected_pattern_distances} and \ref{app:local_bias_mean_field}.

\section{Related literature\label{sec:literature}}

In this section, we present a non-exhaustive overview of the literature related to our work, which lies at the intersection between legal network science, legal complexity, and search models.

\subsection{Legal Network Science}
    In the seminal works \cite{Bommarito2010,Bommarito2011,Katz2014}, the authors apply tools from the area of complex networks to examine the US legal system quantitatively, in particular the structure and content of the US Code and the US Supreme Court citation network. Recent studies focuses on the time-evolution of legal texts in the US and Germany, using a clustering algorithm \cite{Katz2020}; similarly, \cite{Lee2019} correlates changes in the Korean constitutional law to societal changes. Several references within \cite{Katz2020} treat the time-evolution of national and super-national legal corpora from a network-perspective. 
    
    Several authors have noted important analogies between software and legal systems, for instance \cite{Li2015} analyses the US Code based on software engineering terms. More recently, \cite{Coupette2021} has expanded on the analogy, with a focus on symptoms and markers (called \emph{smells}) of software that is likely to become problematic in the future, e.g. through long reference trees or duplicate phrases. 
    
    Further studies base their analysis on topic modelling, a family of machine learning algorithms that extract ``topics'' from a given text (first introduced in \cite{Blei2003}). 
    For instance, \cite{Livermore2010} presents cases studies of the network of SCOTUS opinions, tracking proxies for the change in topic proportions over time. The conceptual bridge to law search is built in \cite{Leibon2018,Dadgosari2020}, by connecting law search and prediction of the relevance of certain documents based on their content and topological properties within the citation network.
    
\subsection{Legal Complexity}

    Some works take a rather system-wide approach, discussing complexity as it relates to the connection between the world, the lawmaker, the legal practitioner, and the citizen that is subject to legal rules. Much of the work in this area is descriptive, such as the influential articles \cite{Schuck1992,White1992,Surden2007}, relating the institutional challenges and ``locations'' of uncertainty in the system. Some authors, e.g. \cite{Sichelman2021, DAmato1983}, make an attempt at general quantification, for instance via the entropy associated to the uncertainty of a legal rule, or by drawing connections between network properties and standards of legal practice \cite{Bourcier2007}.
    
\subsection{(Randomised) Search Models}

    The searching behaviour of agents in a diverse range of systems has been the subject of intensive study for many decades now. Following similar developments in economics \cite{Kohn1974,Diamond1982} and biology \cite{Ramos-Fernandez2004,Boyer2004}, for instance, understanding law searches has been gathering traction recently \cite{Leibon2018,Carlson2020,Dadgosari2020}.

    In these works, searching is framed as an optimal stopping problem: upon sampling a resource (information, trading opportunities, food) a number of times in one location, how does the searcher decide when to stop and change location? Our approach will be different in that our searcher has sufficient information to know exactly when to stop. The modelling of such problems in terms of random walks (on networks or grids, say) has proven successful in many cases -- see for instance \cite{Evans2020} and references therein for search strategies involving resets of the random walker to its starting position. Another similar class of problems concerns moving and hiding targets and optimal strategies for a random-walk searcher \cite{Pandey2018}.

    In \cite{Leibon2018,Carlson2020,Dadgosari2020} the approach is based on a joint empirical analysis of structure and contents of legal text networks, by means of network-based \emph{topic modelling}, a method largely developed in \cite{Blei2003,Blei2007,Blei2012} to extract a set of topics given a sufficiently large body of text, and assign  one or more topic labels to each section. For instance, this analysis can be based on the movements of a random walker in the textual landscape, studying in which regions the walker tends to sojourn for longer periods of time, as well as the overlap between such regions. The authors demonstrate that this analysis may be useful to predict citations in US legal opinions and statutory law. Moreover, the authors propose a law-search model based on their findings on link prediction, and compare it with human law-searchers.  Further references in this area can be found in \cite{Dadgosari2020,Carlson2020}. 
    
    Other lines of research have been more interested in the particular behaviour of an information-seeker. Important foundations to this field are laid out in \cite{Wilson1981}, leading to further studies in various contexts. In particular, \cite{Wilkinson2001} (and references therein) examines the information-seeking behaviour of legal professionals, which relies markedly on informal sources instead of primary literature. Comprehensive reviews of the area are given in \cite{Courtright2007} and \cite{Case2007}.

\section{Definition of the model\label{sec:model}}

We consider a finite tree of $N$ nodes -- in which every node stands for an ``item'' in the law as described in sec.~\ref{sec:intro} -- and two nodes are connected as parent and daughter if one contains the other (e.g. a Section within a Chapter, or an Article within a Section). 
There are no limitations on the exact shape 
of the tree.
For simplicity of the analysis, however, we consider $c$-ary trees, i.e. trees in which a designated \emph{root}, $\rootnode$, has degree $c$ and every other node is either a leaf, or has degree $c+1$.

\begin{figure}[t]
    \centering
    \includegraphics{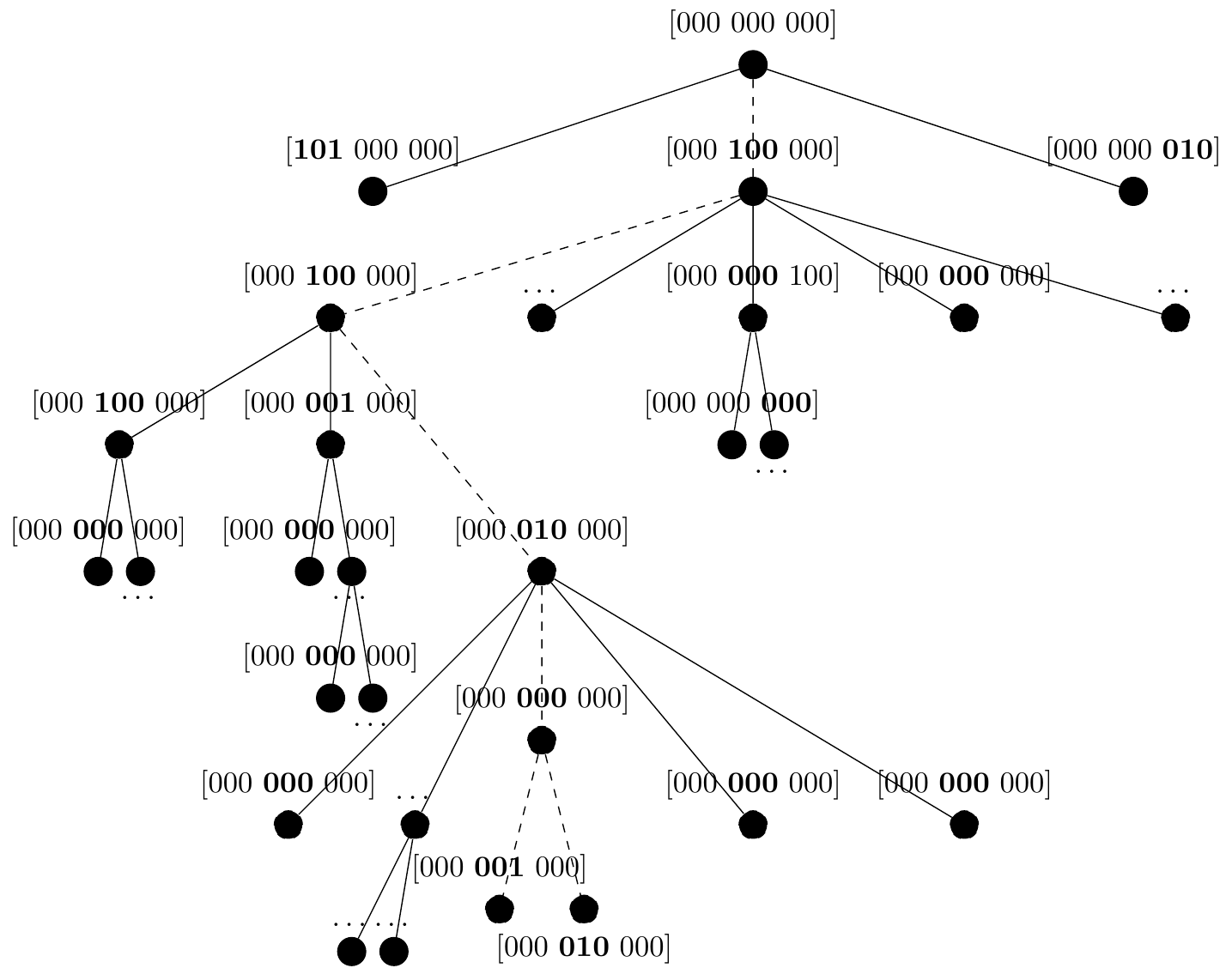}
    \caption{
    Representation of the excerpt of the UK Housing Act 1985 c.68 shown  fig.~\ref{fig:housing_act_example} into binary patterns for the (shortened) glossary $\{ $``slum'', ``demolition'', ``clearance'', ``overcrowding'', ``room'', ``space'', ``responsibilities'', ``occupation'', ``escape''$ \}$ of length $L=9$. Every node is assigned a binary pattern of length $L$, whose bits represent the presence or absence of the corresponding keyword. The boldface bits stand for keywords specific to the Part in which their node is located. Dots are used to omit some of the all-$0$ patterns.
    }
    \label{fig:pattern_demo}
\end{figure}

We model the textual content of every node $v$ in terms of a
binary string of length $L$, that we will refer to as \emph{pattern}.
We denote patterns as
${\bm\xi^v}=(\xi_1^v,\ldots, \xi_L^v)$ where $\xi_i^v\in\{0,1\}$
encodes presence ($1$) or absence ($0$), in the textual content of node $v$, of keyword 
$i$, from a predetermined \emph{glossary} of $L$ keywords (which is typically defined by the user). A reduced glossary for the example 
in fig.~\ref{fig:housing_act_example}
may be the list 
$\{ $``slum'',  ``demolition'', ``clearance'', ``overcrowding'', ``room'', ``space'', ``responsibilities'', ``occupation'', ``escape''$ \}$, which would lead to the assignment of patterns shown in fig.~\ref{fig:pattern_demo}. 

We assume that a reader is interested in the information hidden in a particular leaf of the tree, which we call the \emph{target} $\targetnode$. We model the search process of the reader as a random walker that moves randomly along the links of the tree, starting from the root. 
We assume that the reader is more likely to step on a node 
when the text associated with it has a higher semantic similarity with the sought (target) information. Hence, we assume that when on a node, the walker will step to one of the neighbouring nodes $v$ with a probability that depends on the semantic similarity between the node 
$v$ and the target node, and it does not depend on the starting node. The semantic \emph{dis}similarity of two nodes is measured as the 
\emph{Hamming distance} of their patterns
\begin{equation}\label{eq:hamming}
    d(\bm{\zeta},\bxi):=d_H(\bm{\zeta},\bxi):= \sum_{\ell=1}^L \vert \zeta_\ell - \xi_\ell \vert\ ,
\end{equation}
which equals the number of bits on which the two patterns disagree. We will refer to such distance as ``pattern-distance''. This distance is not to be confused with the ``edge-distance'' between the corresponding nodes on the graph, defined as the number of edges constituting the shortest path between them.
As the probability to step onto a node $v$ does not depend on the node the walker is stepping from, we can define, for any link pointing to $v$, a weight
\begin{equation}\label{eq:edge_weight_function}
\omega(v) = \frac{1}{d\left( \bm\xi^v,\bm\xi^{\targetnode} \right)+1}\ 
\end{equation}
that is higher, the higher the semantic similarity between the patterns in $v$ and $\targetnode$. Using these non-negative weights, we can define a matrix of transition probabilities between two nodes $u$ and $v$ as
\begin{equation}\label{eq:edge_weight}
W_{u,v} = \frac{\omega(v)}{\sum_{x\in\partial u}\omega(x)}\ ,
\end{equation}
where $\partial u$ is the neighbourhood of node $u$.

We will characterise the complexity of the 
search process in terms of the average number of steps taken until the target is first found. 
Our object of study will be the dependence of this quantity on the way patterns are assigned to nodes as well as on the properties of the tree itself.

We will assume a stochastic set-up, where patterns are regarded as quenched random binary vectors, 
with statistics controlled by two tunable parameters, which we call \emph{tightness} and \emph{overlap}, representing the vertical and horizontal coherence of the legal text, respectively. In particular, we will assume that the components of the 
root pattern are 
independent random variables 
with fixed expected value $a$. All patterns on the lower levels
are generated 
from their parent node via a Markov process, 
according to which the entries of the patterns in the child node are mutated with a given rate
with respect to those of the parent node.
The \emph{tightness} $\tau$ is defined as a decreasing function of the mutation rate
such that tighter sets of patterns are generated by least rates of mutation. 
Additionally, assuming that each part covers a unique topic with corresponding specific keywords, we define the \emph{overlap} (denoted $2\Delta$), 
quantifying
the number of keywords that are expected to be
shared by two successive Parts. 

In later sections we will study the impact of $\tau$ and $\Delta$ on the complexity 
of the defined search process. We will first 
quantify their role on the statistics of pattern distances
(sec.~\ref{sec:expected_pattern_distances}), then 
we will study their role on the average number of steps taken by the walker to first reach the target node where the information of interest is hidden  
(secs.~\ref{sec:compfunc} and \ref{sec:simulations}).

In the remainder of this section we 
provide details about the Markov process used 
to generate patterns, which 
reflect the hierarchy of the tree and have the desired properties of tightness and overlap.

\begin{figure}[h]
    \centering
    \includegraphics{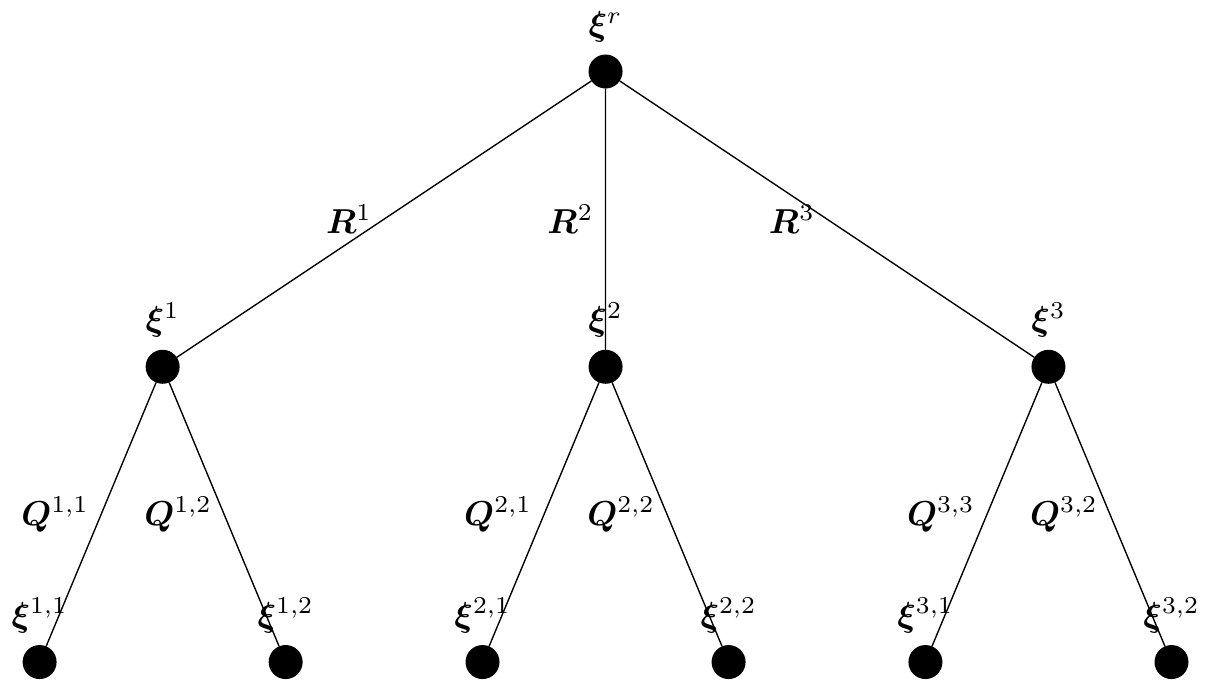}
    \caption{Pattern hierarchy and their relations. The $i$-th  bit of each pattern is generated from the parent bit by a two-state Markov process, the transition matrix of which is indicated next to the edge between the two patterns. For instance, the transition matrix relating $\xi_i^{2}$ to $\xi_i$ is $\bR_i^2$.}
    \label{fig:pattern_Markov_process}
\end{figure}

We will denote the vertices with lexicographical labels, 
such that
the $c$ descendants of the root will be denoted by a single index $\mu_1\in \{1,\ldots,c\}$; the $\mu_2$-th descendant of the $\mu_1$-th descendant of the root, will be denoted by the two indices $\mu_1, \mu_2$ and a node $v$ in the $k$-th generation will be denoted by $k$ indices $\mu_1,\ldots,\mu_k$ where 
$\mu_j\in\{1,\ldots,c\}$ for all $1\leq j\leq k$, see fig.~\ref{fig:pattern_Markov_process} for a schematic representation of the genealogy of patterns. We 
denote with 
$\bxi$
the textual pattern at the root. 
For the root pattern, we assume the entries to be drawn from a factorised distribution
\begin{equation}
    P(\bxi)=\prod_{i=1}^L P_{i}(\xi_i)
\end{equation}
with 
\begin{equation}\label{eq:a_def}
    P_i(\xi_i)=a\xi_i+(1-a)(1-\xi_i)\ ,
\end{equation}
so for all $i=1,\dots,L$ the expectation is $\anglemean{\xi_i}=a $.
For the first level of hierarchy (i.e. Part-level), we assume that the patterns are generated with distribution
\begin{equation}
    P^\mu(\bxi^\mu \vert \bxi)= \prod_{i=1}^L R_{i}^\mu(\xi_i^\mu \vert \xi_i )
\end{equation}
from the root pattern. Here, the $R_{i}^\mu(\xi_i^\mu \vert \xi_i)$ for $\xi_i^\mu,\xi_i \in \{0,1\}$ form the entries of the $2\times 2$ transition matrix of the Markov process generating the Part-pattern entry $\xi_i^\mu$ from the root. By the law of total probability, $\xi_i^\mu$ obeys the marginal probability
\begin{align}
    P_i^\mu(\xi_i^\mu)=\sum_{\xi_i =0}^1 P_i ( \xi_i )R_{i}^{\mu}(\xi_i^\mu \vert \xi_i )\ .
\end{align}
Fixing $\anglemean{\xi_i^\mu}=a_i^\mu$, 
$\bR_{i}^{\mu}$ has the elements
\begin{equation}\label{eq:R_def}
    \bR_{i}^{\mu} =
    \begin{pmatrix}
        P_i^\mu(0 \vert 0 ) & P_i^\mu(1 \vert 0) \\
        P_i^\mu(0 \vert 1 ) & P_i^\mu(1 \vert 1)
    \end{pmatrix}
    =
    \begin{pmatrix}
        1-\Gammap & \Gammap \\
        1-\frac{a^\mu_i -(1-a)\Gammap}{a} & \frac{a^\mu_i -(1-a)\Gammap}{a}
    \end{pmatrix}\ ,
\end{equation}
where the parameter $\Gammap\in [0,1]$ determines the rate of mutation from $\xi_i$ to $\xi_i^\mu$.

\begin{figure}[h]
    \centering
    \includegraphics{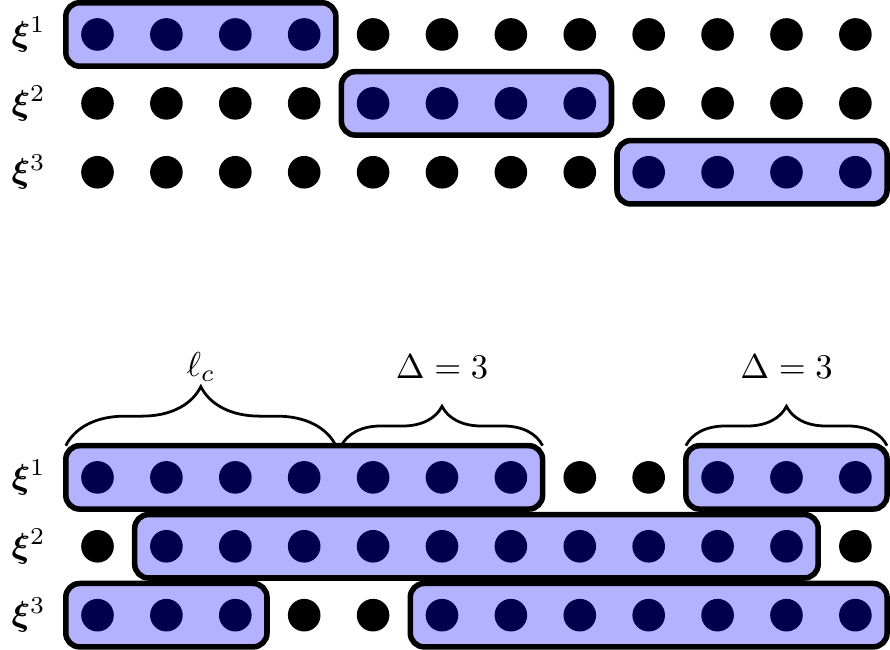}
    \caption{A schematic drawing of patterns with $L=12$ and $c=3$, with boxes marking bits with high expectation $\anglemean{\xi_i^\mu}=\ah$. Top: $\Delta=0$, i.e. specific keywords of different Parts do not overlap. Bottom: $\Delta=3$, i.e. two neighbouring Parts have $2\Delta=6$ specific keywords in common.
    For $\Delta=0$ there are $\lc (c-1)$ keywords that are generic for any Part, therefore the maximum value for $\Delta$ is $\lc \frac{(c-1)}{2}=4$. }
    \label{fig:overlap_schema}
\end{figure}

We take the $a_i^\mu$ to satisfy some constraints,
motivated by the idea that different Parts treat individual topics. If a given keyword is highly related to the topic of some Part, it will have a high probability to appear in that Part. If each keyword is related to one topic only, it will appear with high probability in the Part treating that topic, and with low probability in all other Parts. This situation is represented in the top panel of fig.~\ref{fig:overlap_schema}. However, we allow for a degree of topic similarity between two successive Parts $\mu$ and $\mu+1$, realised by a subset of size $2\Delta$ of keywords that appear with high probability in $\mu$, as well as in $\mu+1$. This situation for $2\Delta=6$ is shown in the bottom panel of fig.~\ref{fig:overlap_schema},
representing the appearance of topic-specific keywords in the Parts. In particular, we 
introduce 
$\Delta$ as a parameter that 
controls the number of Part-specific keywords shared by neighbouring Parts $\mu, \mu+1$
i.e. as the Part \emph{overlap}, and define

\begin{equation}\label{eq:ah_al_def}
    a^\mu_i = \begin{cases}
        \ah \quad &\colon \ (\mu-1) \lc-\Delta < i \leq \mu\lc+\Delta\ , \\
        \al \quad &\colon \ \textup{else}\ ,
    \end{cases}
\end{equation}
where $\lc$ is the ratio $L/c$, $\Delta\in [0,\lc\frac{c-1}{2}]$ 
and we consider
periodic boundaries, i.e. $a^\mu_i=a^\mu_{kL+i}$ for all $k\in\mathbb{Z}$ (see fig.~\ref{fig:overlap_schema} for a schematic representation). Moreover, we assume $0< a_l<a_h <1$ (the indices $l$ and $h$ stand for ``low'' and ``high'').
However, since elements of $\bR_{i}^{\mu}$ are probabilities, one can derive the stricter bounds
\begin{equation}\label{eq:marginal_boundaries}
    (1-a)\Gammap \leq a^\mu_i \leq (1-a)\Gammap+a\ 
\end{equation}
for all $\mu=1,\dots, c$ and $j=1,\dots, L$.
For our purposes it will be useful to define
\begin{align}\label{eq:al_betal}
    \al &= (1-a)\Gamma^\prime + \betal\\
    \label{eq:ah_betah}
    \ah &= (1-a)\Gamma^\prime + a
\end{align}
for some $\beta_l$ satisfying $0 < \betal <a $ to enforce $\al < \ah$.
The set of indices $i$ such that $a^{\mu}_i=\ah$ we refer to as the \emph{$\ah$-domain} of $\mu$, and to the complementary set as \emph{$\al$-domain}.

We now extend the above prescription to 
generate patterns at the lower levels of the hierarchy. 
Consider the pattern $\bxi^{\mu_1,\dots \mu_k}$ on level $k$, generated with probabilities 
\begin{equation}
    P^{\mu_1 \dots \mu_k}(\bxi^{\mu_1 \dots \mu_k} \vert \bxi^{\mu_1\dots\mu_{k-1}}) = \prod_{i=1}^L Q_{i}^{\mu_1\dots \mu_k}(\xi_i^{\mu_1 \dots \mu_k} \vert \xi_i^{\mu_1 \dots \mu_{k-1}})
\end{equation}
for a $2\times 2$ transition matrix $\bQ_{i}^{\mu_1\dots \mu_k}$,
so the marginal probabilities can be written in terms of the marginal of the parent pattern
\begin{equation}
    P_{i}^{\mu_1 \dots \mu_k}(\xi_i^{\mu_1 \dots \mu_k})=\sum_{\xi_i^{\mu_1\dots\mu_{k-1}}=0}^1 P_i^{\mu_1\dots\mu_{k-1}}(\xi_i^{\mu_1\dots\mu_{k-1}})Q_{i}^{\mu_1\dots \mu_k}(\xi_i^{\mu_1 \dots \mu_k} \vert \xi_i^{\mu_1 \dots \mu_{k-1}})\ .
\end{equation}
For simplicity, we assume that $a_i^{\mu_1\dots \mu_k}:= \anglemean{ \xi_i^{\mu_1 \dots \mu_k} }$ is inherited from the parent, i.e. 
\begin{equation}\label{eq:identical_expectations}
    a^{\mu_1\dots,\mu_{k}}_i = a^{\mu_{1}\dots\mu_{k-1}}_i=\dots=a^{\mu_1}_i\ ,
\end{equation}
which equals either $a_h$ or $a_l$, according to eq.~\eqref{eq:ah_al_def}.
Then $\bQ_{i}^{\mu_1 \dots \mu_k}$ takes the form
\begin{equation}\label{eq:Q_def}
    \bQ_{i}^{\mu_1\dots \mu_k}=
    \begin{pmatrix}
        P_{i}^{\mu_1\dots\mu_k}(0\vert 0) & P_{i}^{\mu_1\dots\mu_k}( 1 \vert 0 ) \\
        P_{i}^{\mu_1 \dots \mu_k}(0 \vert 1 ) & P_{i}^{\mu_1 \dots \mu_k}( 1 \vert 1 )
    \end{pmatrix} \\
    =\begin{pmatrix}
        1-a^{\mu_1}_i\Gamma & a^{\mu_1}_i\Gamma \\
        (1-a_i^{\mu_1})\Gamma & 1-(1-a_i^{\mu_1})\Gamma
    \end{pmatrix}\ .
\end{equation}
Here the parameter $\Gamma\in [0,1]$ controls the level of noise in the patterns below Part-level.
The relation between patterns in terms of the transition matrix families $\bR$ and $\bQ$ is summarised in fig.~\ref{fig:pattern_Markov_process}.

With all model parameters defined, in the next section we 
study the statistical properties of the distances 
between patterns sitting on different nodes of the network, as these will determine the kinetics of the random walk and, in particular, the complexity of the search process that the random walk is meant to model.

\section{Expected Pattern-Distances Along and Across Branches}\label{sec:expected_pattern_distances}
In this section 
we provide analytical expressions, in terms of the model control parameters $\tau$ and $\Delta$, 
for
the expected values of two classes of pattern-distances, namely between
(1) adjacent Part-level patterns $\bxi^\mu$ and $\bxi^{\mu+1}$ and
(2) any Part-level pattern $\bxi^{\mu_1}$ and leaf patterns of the same Part, $\bxi^{\mu_1\dots\mu_h}$.
For clarity of presentation, we state here the main results and present their derivations in app.~\ref{app:expected_pattern_distances}.

\subsection{Overlap: Distance between neighbouring patterns}\label{sec:overlap_expected_distance}

Let $\bxi^{\mu}$, $\bxi^{\mu+1}$ be two child patterns of the root pattern $\bxi$, with marginal expectations as described by eq.~\eqref{eq:ah_al_def}.
With $d$ being the Hamming distance, eq.~\eqref{eq:hamming}, we are interested in the properties of the distance on Part-level $d^{\mu,\mu+1}:=d\left( \bxi^\mu,\bxi^{\mu+1} \right)$ as we vary $\Delta$. App.~\ref{app:expected_pattern_distances} shows that 
the 
expectation of the
pattern-distance 
over the distribution of patterns
$\bad(\Delta):= \anglemean{ d^{\mu,\mu+1}  }  $ 
is given by the expression
\begin{equation}\label{eq:overlap_expected_distance}
   \bad(\Delta)= \begin{cases}
        \bad_0 + 2\lc(a-\betal)\left((c-2)\frac{\betal}{a}+1 \right) -4\Delta \frac{(a-\betal)\betal}{a} \quad &\colon \Delta \leq \Deltatrans\ , \\ 
        \bad_0 + 2\lc(c-1)(a-\betal) - 4\Delta(a-\betal) \quad &\colon \ \textup{else}\ ,
    \end{cases}
\end{equation}
with $\Deltatrans=\lc \frac{(c-2)}{2}$ and $\bad_0=2(1-a)\Gammap(1-\Gammap)L$. We recall that $\lc=L/c$, where
$L$ and $c$ are the pattern length and the number of children of each node, respectively.

\begin{figure}[h]
    \centering
    \includegraphics{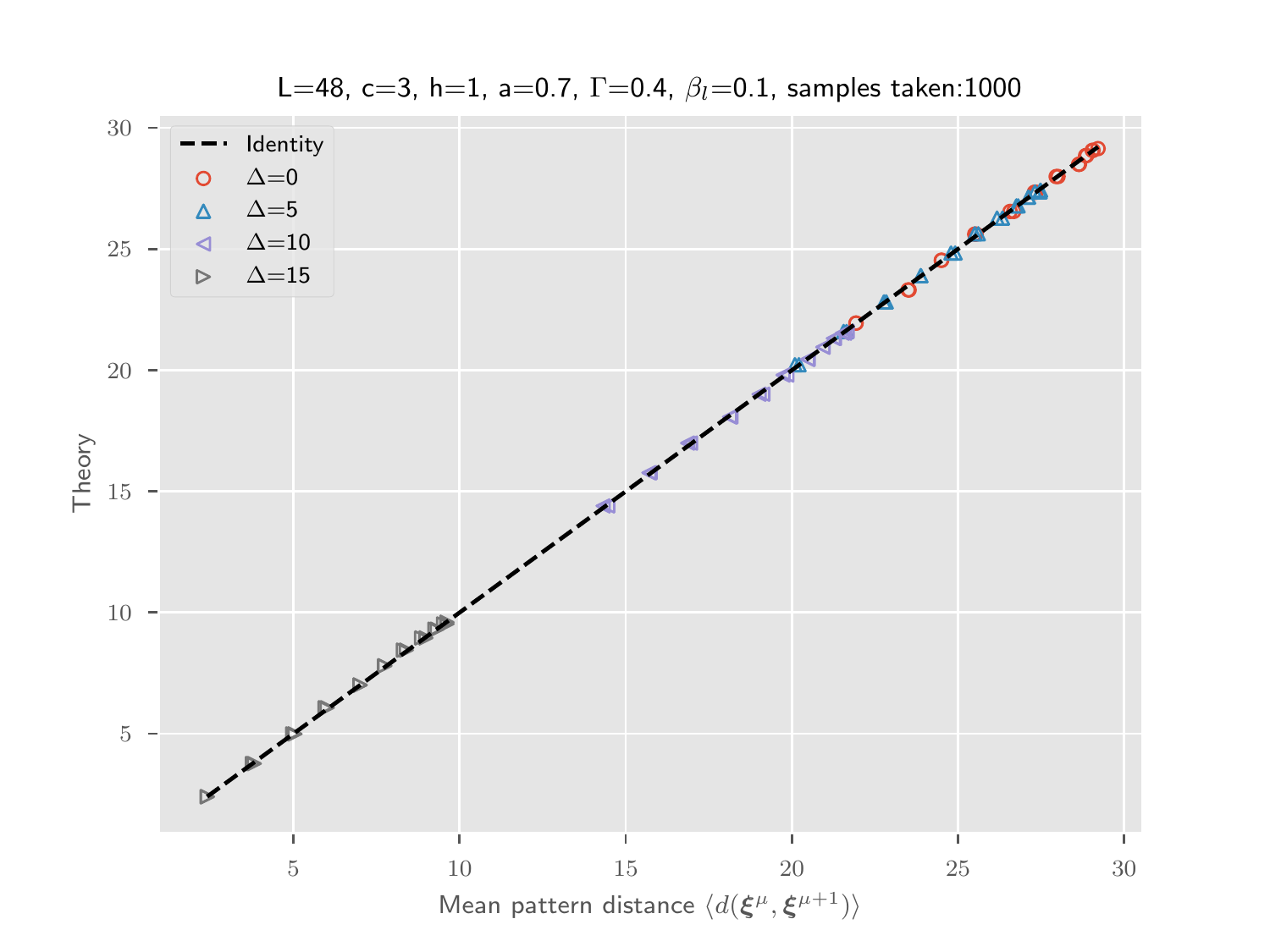}
    \caption{Scatter plot for the expected distance $\bad$ on the horizontal axis, compared to eq.~\eqref{eq:overlap_expected_distance} on the  vertical axis. $\bad$ is sampled in simulations with the parameters shown at the top.}
    \label{fig:overlap_expected_distance}
\end{figure}

Fig.~\ref{fig:overlap_expected_distance} compares eq.~\eqref{eq:overlap_expected_distance} to the numerical average of distances $d^{\mu,\mu+1}$. The agreement is excellent, showing the accuracy of our calculation.

We observe that $\bad$ is a decreasing function of $\Delta$, in the physical range of parameters 
$\betal<a$ that we identified after  
eq.~\eqref{eq:al_betal}. 
Hence, the parameter $\Delta$ controls the ``topical overlap'' between adjacent Parts as described in the previous section.

\subsection{Tightness: Distance between ancestor and descendant patterns}\label{sec:tightness_expected_distance}

\begin{figure}[h]
    \centering
    \includegraphics{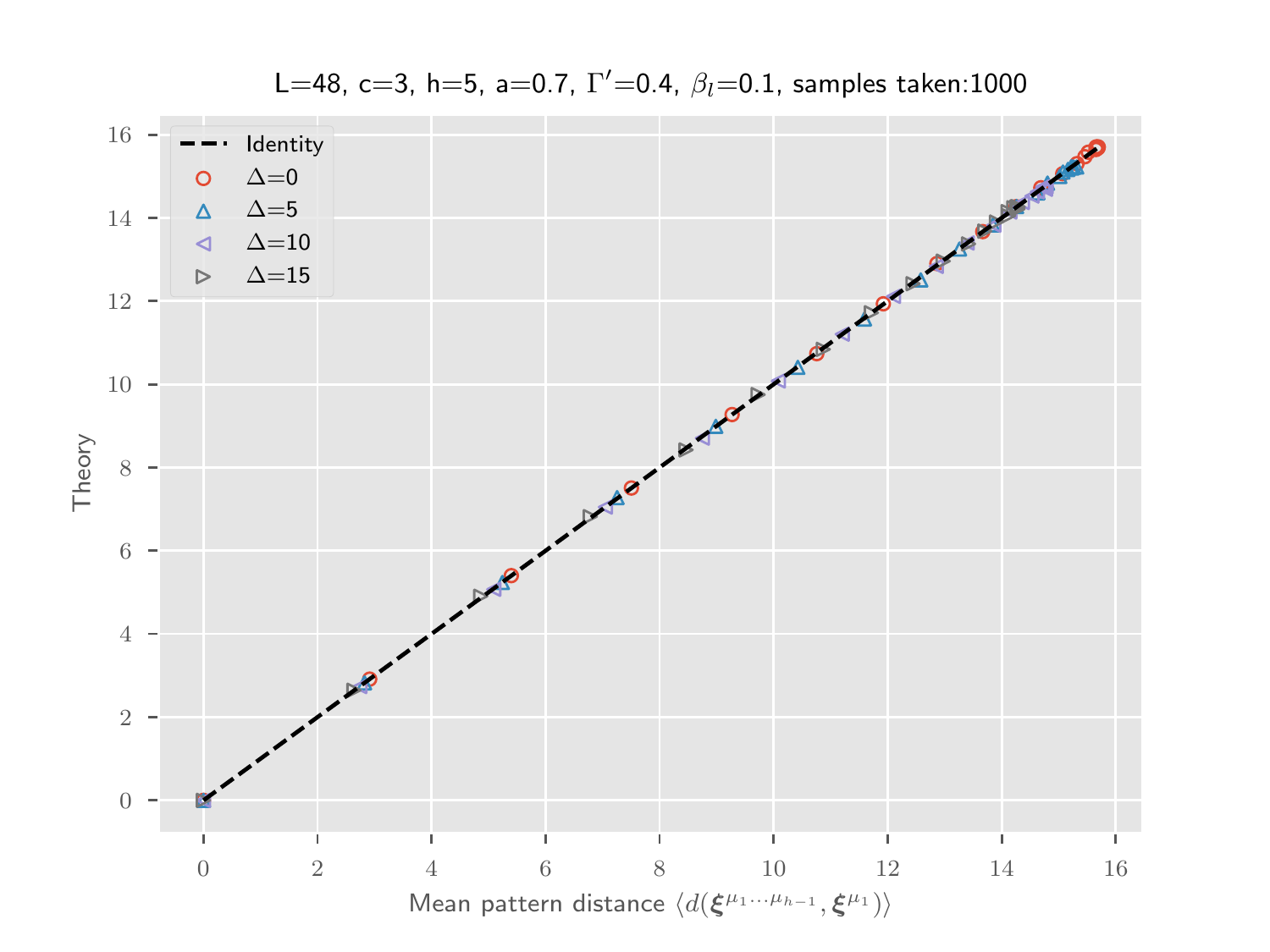}
    \caption{Mean pattern-distance across $h=4$ levels, simulated with the shown parameters on the $x$-axis, and the predicted expectation value according to eq.~\eqref{eq:tightness_expected_distance} on the $y$-axis.}
    \label{fig:tightness_expected_distance}
\end{figure}

Having examined the ``horizontal'' variation of patterns in the previous section, we now turn to ``vertical'' pattern-distance; that is to say, the expected pattern-distance between the ``Part'' node at the top of a certain branch, and any leaf in the same branch. To this end, let $\bxi^{\mu_1}$ be any Part-level pattern, and let $\bxi^{\mu_1\dots \mu_{h-1}}$ be the pattern of any leaf in the same branch, i.e. descending from ${\mu_1}$. Appendix~\ref{app:expected_pattern_distances} shows that we have for the expected pattern-distance between $\bxi^{\mu_1}$ and $\bxi^{\mu_1\dots\mu_{h-1}}$
\begin{equation}\label{eq:tightness_expected_distance}
    \anglemean{ d(\bxi^{\mu_1} , \bxi^{\mu_1\dots\mu_{h-1}} ) } = g^{(h-1)}(\al) \left(\lc (c-1)-2\Delta\right) + g^{(h-1)}(\ah) \left(\lc+2\Delta\right)\ ,
\end{equation}
with 
\begin{equation}
    g^{(k)}(x):=1-2x(1-x)\left( 1-\Gamma \right)^{k}\ . 
\end{equation}
In fig.~\ref{fig:tightness_expected_distance} we compare eq.~\eqref{eq:tightness_expected_distance} to the numerical average of $d(\bxi^{\mu_1} , \bxi^{\mu_1\dots\mu_{h-1}} )$, observing an excellent agreement between the two.

In contrast to $\Delta$, the corresponding pattern-distance does not depend linearly on $\Gamma$. However, as it strictly decreases as $\Gamma$ increases, we see that 
$\Gamma$ acts as expected from a mutation rate, namely, 
the higher the mutation rate, the higher
the distance between Part- and leaf-level patterns.
Eq.~\eqref{eq:Q_def} shows that upon setting $\Gamma=1$, the $\xi_i^{\mu_1}$'s and $\xi_i^{\mu_1\dots \mu_{h-1}}$'s become independent. This is the state of least tightness and also the maximum of $\anglemean{ d(\bxi^{\mu_1} , \bxi^{\mu_1\dots\mu_{h-1}} ) }$. Similarly, $\Gamma=0$ enforces $\anglemean{ d(\bxi^{\mu_1} , \bxi^{\mu_1\dots\mu_{h-1}} ) }=0$, representing highest tightness. 
We define the \emph{tightness} as the following 
monotonically decreasing function of $\Gamma$
\begin{equation}\label{eq:tightness_Gamma}
    \tau(\Gamma)=1-g^{(h-1)}(a_j)/[2a_j(1-a_j)] = \left(  1- \Gamma \right)^{h-1}\ .
\end{equation}
 As discussed in sec.~\ref{sec:model}, this parameter controls the (expected) similarity between an ancestor-descendant pair of patterns, just as $\Gamma$ controls the level of mutation from the former to the latter. 
 Tuning $\tau$ (as opposed to $\Gamma$) allows us to 
 achieve a better resolution in simulations for $\Gamma\to 1$.

In analogy to the relation between $\Delta$ and $\bad$, the distance $\anglemean{ d(\bxi^{\mu_1} , \bxi^{\mu_1\dots\mu_{h-1}} ) }$ is
linearly decreasing in $\tau$.

\section{Brief Review of Mean First-Passage Times\label{sec:mfpt_rev}}

The mean first-passage time (MFPT) $m_{ij}$ of a random walker is the expected time the walker takes to first hit node $j$, having started from node $i$; the expectation is taken over many random walks on a fixed instance of the graph.
MFPTs can be derived as unique solutions of the recurrence equations
\begin{equation}\label{eq:mfpt_basic_recurrence}
    m_{ij} = W_{ij}+\sum_{k\neq j} W_{ik}(m_{kj}+1)\ ,
\end{equation}
where $W_{ik}$ denotes the transition probability of the walker from node $i$ to node $k$.
The first term in eq.~\eqref{eq:mfpt_basic_recurrence} accounts for the walker hopping from $i$ to $j$ directly (which occurs with probability $W_{ij}$), while the second term accounts for the walker hopping to any other node $k$ first and 
starting a first-passage process from there (at the next time step).

Eq.~\eqref{eq:mfpt_basic_recurrence} can be rearranged into \cite{Masuda2017}
\begin{equation}\label{eq:mfpt_grounded_laplacian}
    \bm{m}_j = \left( \Eye_{N-1} - \widehat{\bm{W}}_j \right)^{-1}\bm{1}_{N-1},
\end{equation}
where $\Eye_{N-1}$ and $\bm{1}_{N-1}$ are the unit matrix and the  all-one vector of size $N-1$, respectively, and $\bm{m}_j$ is the vector of MFPTs to node $j$, whose element $(\bm{m}_j)_i$ corresponds to the MFPT from source node $i$ for $i \in \{ 1,\dots,N \}\setminus\{ j \}$ to $j$. Moreover, in eq.~\eqref{eq:mfpt_grounded_laplacian}, $\widehat{\bm{W}}_j$ is the matrix obtained by removing the $j$-th row and column from $\bW$. In this paper,  eq.~\eqref{eq:mfpt_grounded_laplacian} will be used to provide a numerical benchmark to compare our analytical estimates with. Further information on methods and applications of MFPTs can be found in \cite{Aldous1999} and the excellent review \cite{Masuda2017} (as well as references in the latter).

\begin{figure}[h]
    \centering
    \includegraphics{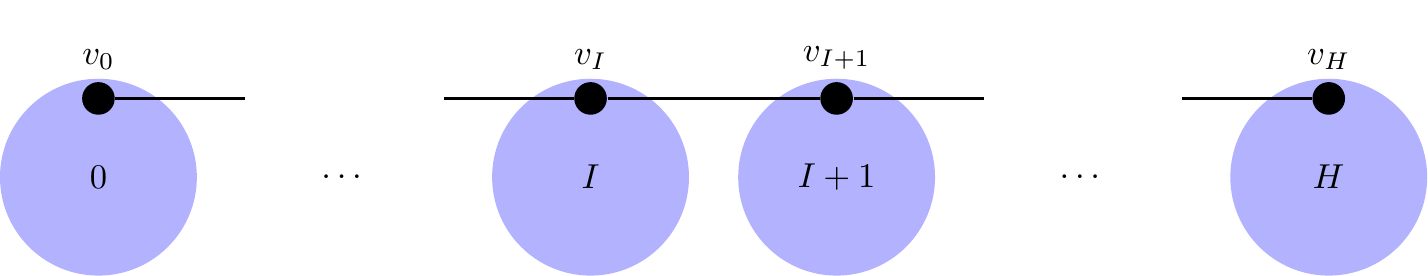}
    \caption{A network of clusters $0, 1, \dots, H$ lined up on the nodes $v_0,v_1,\dots, v_H$.}
    \label{fig:necklace}
\end{figure}
\begin{figure}[h]
    \centering
    \includegraphics{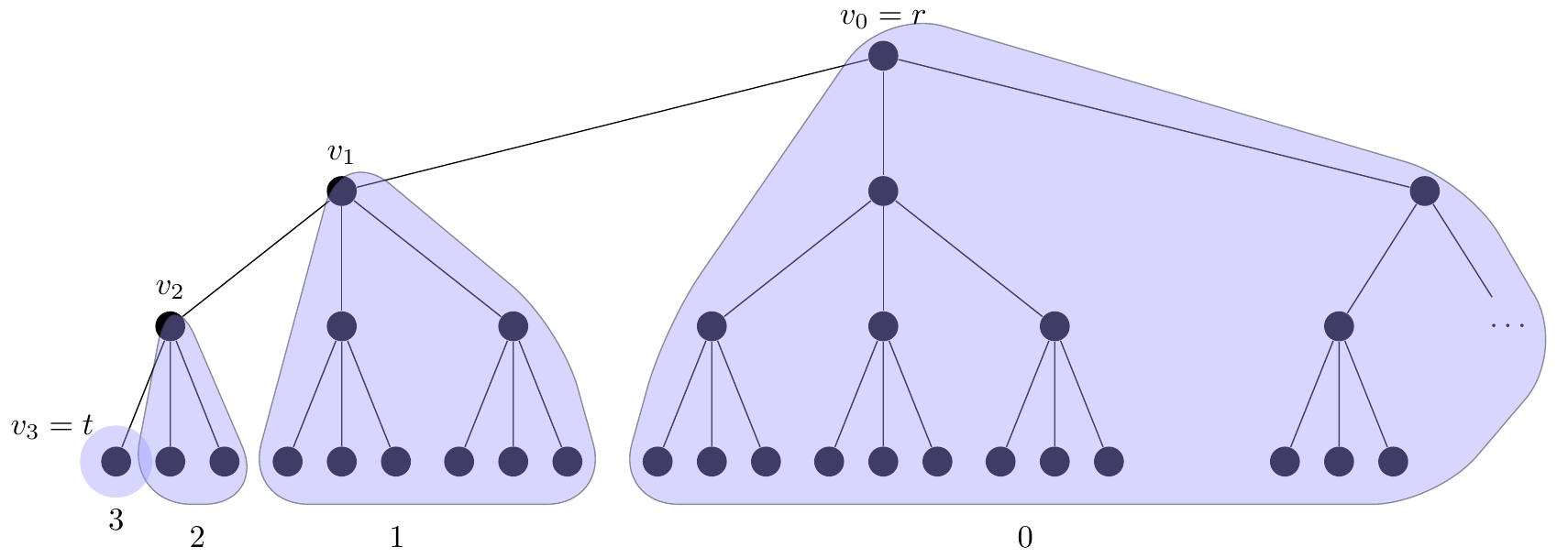}
    \caption{Ternary tree with height $h=3$. Shaded areas show clusters $0,1,2,3$.}
    \label{fig:tree_clustering}
\end{figure}

For
tree-graphs, which we consider in this manuscript, 
and generally for graphs in which two sets of nodes exist that are connected by a single edge, one can find more explicit formulae for MFPTs. As we have shown elsewhere \cite{Forster2022}, if we can coarse-grain the network into clusters $0,\dots,H$ that lie on a line, and every cluster $K$ hangs from a one-dimensional chain only via a single node $v_K$ as shown in fig.~\ref{fig:necklace}, the MFPT $m_{v_0,v_H}$ is given by the formula
\begin{equation}\label{eq:mfpt_necklace}
    m_{v_0,v_H} = \sum_{I=1}^{H}  \sum_{K=0}^{I-1} \frac{W_{v_{I-1},v_{I-2}}\cdots W_{v_{K+1} v_K}}{W_{v_{K},v_{K+1}} \cdots W_{v_{I-1},v_{I}}} \frac{\Pi_{K}}{\pi_{v_K}}\ ,
\end{equation}
where $\bpi^T$ is the stationary probability vector of the transition matrix $W$ -- i.e. $\bpi^T$ is the unique solution of the eigenvector equation $\bpi^T\bm{W}=\bpi$ -- and $\Pi_K = \sum_{k\in K}\pi_{k} $ is the stationary probability of the cluster $K$. Fig.~\ref{fig:tree_clustering} exemplifies how the clusters can be defined if the network is a tree.

We finish this review with a useful note on how $\bpi^T$ can be derived explicitly without solving the eigenvector problem $\bpi^T\bm{W}=\bpi$ directly if the graph is a tree. 
Let $u$ be any node, and let $\frakt_u$ be the collection of directed edges pointing towards $u$. 
A directed edge $(v \to w)$ is \emph{pointing towards} $u$ if the distance between $w$ and $u$ is less than the distance between $v$ and $u$. In that case, $(v \to w) \in \frakt_u$, and otherwise $(w \to v)\in \frakt_u$, noticing that one of the two has to be the case if the full graph is a tree.

It can then be shown that $\pi_u$ is proportional to the product of the weights $W_{v w}$ running over $\frakt_u$ (see e.g. \cite{Pitman2018}),
\begin{equation}\label{eq:pi_spanning_trees}
    \pi_{u}=\frac{ \prod_{(v\to w)\in \frakt_u } W_{v w} }{ \sum_{s} \prod_{(v \to w)\in \frakt_s } W_{v w} }\ .
\end{equation}
Here, the sum in the denominator runs over all nodes of the graph to ensure that $\bpi^T$ is normalised. 

\begin{figure}[h]
    \centering
    \includegraphics{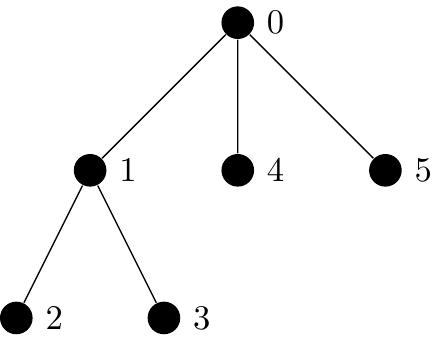}
    \caption{A tree on six vertices to demonstrate the use of eq.~\eqref{eq:pi_spanning_trees}.}
    \label{fig:tree_formula_example}
\end{figure}

As an application of eq.~\eqref{eq:pi_spanning_trees}, consider a random walk on the tree
in fig.~\ref{fig:tree_formula_example} with a general transition matrix $\bW$. To find $\bpi^T$, we have to solve the linear system of six equations given by
\begin{align}
    \bm{x}^T\bW=\bm{x}^T\ ,
\end{align}
which is elementary but laborious. Instead, we can apply eq.~\eqref{eq:pi_spanning_trees}, which tells that $\pi_{0}$, say, is proportional to the product
\begin{align}
    \pi_{0}=\frac{1}{Z} W_{1,0}W_{4,0}W_{5,0}W_{2,1}W_{3,1}\ .
\end{align}
Analogously, the other entries of $\bpi^T$ are proportional to
\begin{align}
    \nonumber
    \pi_{1} &=\frac{1}{Z} W_{0,1}W_{4,0}W_{5,0}W_{2,1}W_{3,1}\ , \\ \nonumber
    \pi_{2} &=\frac{1}{Z} W_{0,1}W_{4,0}W_{5,0}W_{1,2}W_{3,1}\ , \\ \nonumber
    \pi_{3} &=\frac{1}{Z} W_{0,1}W_{4,0}W_{5,0}W_{2,1}W_{1,3}\ , \\ \nonumber
    \pi_{4} &=\frac{1}{Z} W_{1,0}W_{0,4}W_{5,0}W_{2,1}W_{3,1}\ , \\
    \pi_{5} &=\frac{1}{Z} W_{1,0}W_{4,0}W_{0,5}W_{2,1}W_{3,1}\ , 
\end{align}
where the prefactor $\frac{1}{Z}$ ensures that $\bpi$ is normalised.

\section{Complexity Measure for Legal Trees from Approximate MFPT\lowercase{s}\label{sec:compfunc}}

In this section, we present an expression for the \emph{complexity} of the Act represented by the model introduced in sec.~\ref{sec:model}. 
For the definition of the complexity, recall that every assignment of patterns defines a transition matrix $\bW$ for the random walker due to eq.~\eqref{eq:edge_weight}. Denoting by $\mrt(\bW)$ the MFPT from root $\rootnode$ to target $\targetnode$ given such a transition matrix, we define the complexity as the average of $\mrt(\bW)$ over the distribution of patterns,
\begin{equation}\label{eq:complexity_def}
	C:=\anglemean{\mrt(\bW)}\ .
\end{equation}
The quantity $C$ does not depend on any particular realisation of patterns; it reflects the ``higher-level'' properties encoded in the model parameters and the tree. However, evaluating the expectation in eq.~\eqref{eq:complexity_def}
based on the formulae in eq.~\eqref{eq:mfpt_grounded_laplacian} or eq.~\eqref{eq:mfpt_necklace} analytically is a formidable task. We avoid this difficulty by introducing the mean-field approximation
\begin{align}\label{eq:complexity_MF_def}
	\CMF := \mrt \left( \anglemean{\bW} \right)  \approx  \anglemean{ \mrt(\bW) }=C \ ,
\end{align}
i.e. we calculate $\mrt$ for the random walker subject to the averaged  transition matrix $\anglemean{\bW}$, with the average taken over the pattern distribution. Since $\bW$ is always a stochastic matrix, so is $\anglemean{\bW}$, and it does indeed define a random walker on the tree.
Eq.~\eqref{eq:complexity_MF_def} is an approximation because $\mrt$ is a non-linear function of $\bW$, which can be seen from eq.~\eqref{eq:mfpt_grounded_laplacian}.
We dub the approximate, left-hand quantity in eq.~\eqref{eq:complexity_MF_def} the \emph{approximate complexity}.

$\CMF$ is an explicit (though complicated) function of all model parameters, though we are mostly interested
its dependencies on $\Delta$ (defined in eq.~\eqref{eq:ah_al_def}), $\tau$ (defined in eq.~\eqref{eq:tightness_Gamma}) and $a$ (defined in eq.~\eqref{eq:a_def}).
Furthermore, recall the parameters $h$ and $c$ of the tree itself, being the height of the tree, and the number of children to a (non-leaf) node, respectively.
The derivation of $\CMF$ is deferred to app.~\ref{app:MFPT_mean_field}; 
in the following we only summarise the results.
Due to the cumbersome nature of the expressions, they will also be made available as Python code. 

$\CMF$ decomposes into $h$ summands
\begin{align}\label{eq:complexity_full}
    \CMF={\CMF}_0+{\CMF}_1+\sum_{K=2}^{h-1}\ {\CMF}_K\ .
\end{align}
The constituents are given by
\begin{align}    \nonumber
    \nonumber
    {\CMF}_0=& \frac{\Pi_0}{\pi_{v_0}}\frac{\prod_{\mu=2}^c\varepsilon\left( f_\Delta^{(h-1,0,0,0;1)},f_\Delta^{(h-1,1,1,0;\mu)} \right)+\sum_{\mu=2}^c\prod_{\nu=2;\nu\neq\mu}^{c}\varepsilon\left( f_\Delta^{(h-1,0,0,0;1)},f_\Delta^{(h-1,1,1,0;\nu)} \right) }{\varepsilon\left( f_\Delta^{(h-2,0,0,0;1)},f_\Delta^{(h-1,1,0,0;1)}\right)\varepsilon\left( f_\Delta^{(h-2,0,0,0;1)},f_\Delta^{(h,0,0,0;1)}\right)\prod_{\mu=2}^c\varepsilon\left( f_\Delta^{(h-1,0,0,0;1)},f_\Delta^{(h-1,1,1,0;\mu)} \right)}\\ 
            &\times \sum_{I=1}^{h-1}\prod_{J=2}^{I-1}\frac{1}{\varepsilon\left( f_\Delta^{(h-J-1,0,0,0;1)},f_\Delta^{(h-J+1,0,0,0;1)} \right)} \\ \nonumber
     {\CMF}_1=& \frac{\Pi_1}{\pi_{v_1}}\frac{1}{\varepsilon\left( f_\Delta^{(h-2,0,0,0;1)},f_\Delta^{(h-1,1,0,0;1)} \right)\varepsilon\left( f_\Delta^{(h-2,0,0,0;1)},f_\Delta^{(h,0,0,0;1)} \right)}\\ \nonumber
    &\times \left( (c-1)\varepsilon\left( f_\Delta^{(h-2,0,0,0;1)},f_\Delta^{(h-1,1,0,0;1)} \right)+\varepsilon\left( f_\Delta^{(h-2,0,0,0;1)},f_\Delta^{(h,0,0,0;1)} \right)\right. \\ \nonumber
    &\quad +\left. \varepsilon\left( f_\Delta^{(h-2,0,0,0;1)},f_\Delta^{(h-1,1,0,0;1)} \right)\varepsilon\left( f_\Delta^{(h-2,0,0,0;1)},f_\Delta^{(h,0,0,0;1)} \right)  \right)
    \\ 
    &\times \sum_{I=2}^{h-1}\prod_{J=2}^{I-1}\frac{1}{\varepsilon\left( f_\Delta^{(h-J-1,0,0,0;1)},f_\Delta^{(h-J+1,0,0,0;1)} \right)} \\ \nonumber
    {\CMF}_K=&  \frac{\Pi_K}{\pi_{v_K}} \left[c+\varepsilon\left( f_\Delta^{(h-K-1,0,0,0;1)},f_\Delta^{(h-K+1,0,0,0;1)} \right) \right]\\ 
        &\times \sum_{I=K+1}^{I-1}\prod_{J=K}^{I-1}\frac{1}{\varepsilon\left( f_\Delta^{(h-J-1,0,0,0;1)},f_\Delta^{(h-J+1,0,0,0;1)} \right)}\ ,
\end{align}
where the fractions $\Pi_K/\pi_{v_K}$ for $K>1$, $\Pi_1/\pi_1$ and $\Pi_0/\pi_0$ are given in app.~\ref{app:MFPT_mean_field} in eqs.~\eqref{eq:PiJ/piJ}, \eqref{eq:Pi1/pi1} and \eqref{eq:Pi0/pi0}, respectively.
Moreover, the weight functions $\varepsilon$ are defined in eq.~\eqref{eq:epsilon_def}, and the $f_\Delta$'s in eqs.~\eqref{eq:dv_binomial_approx} and \eqref{eq:switching_rates_explicit}.

\begin{figure}[h]
	\centering
	\includegraphics{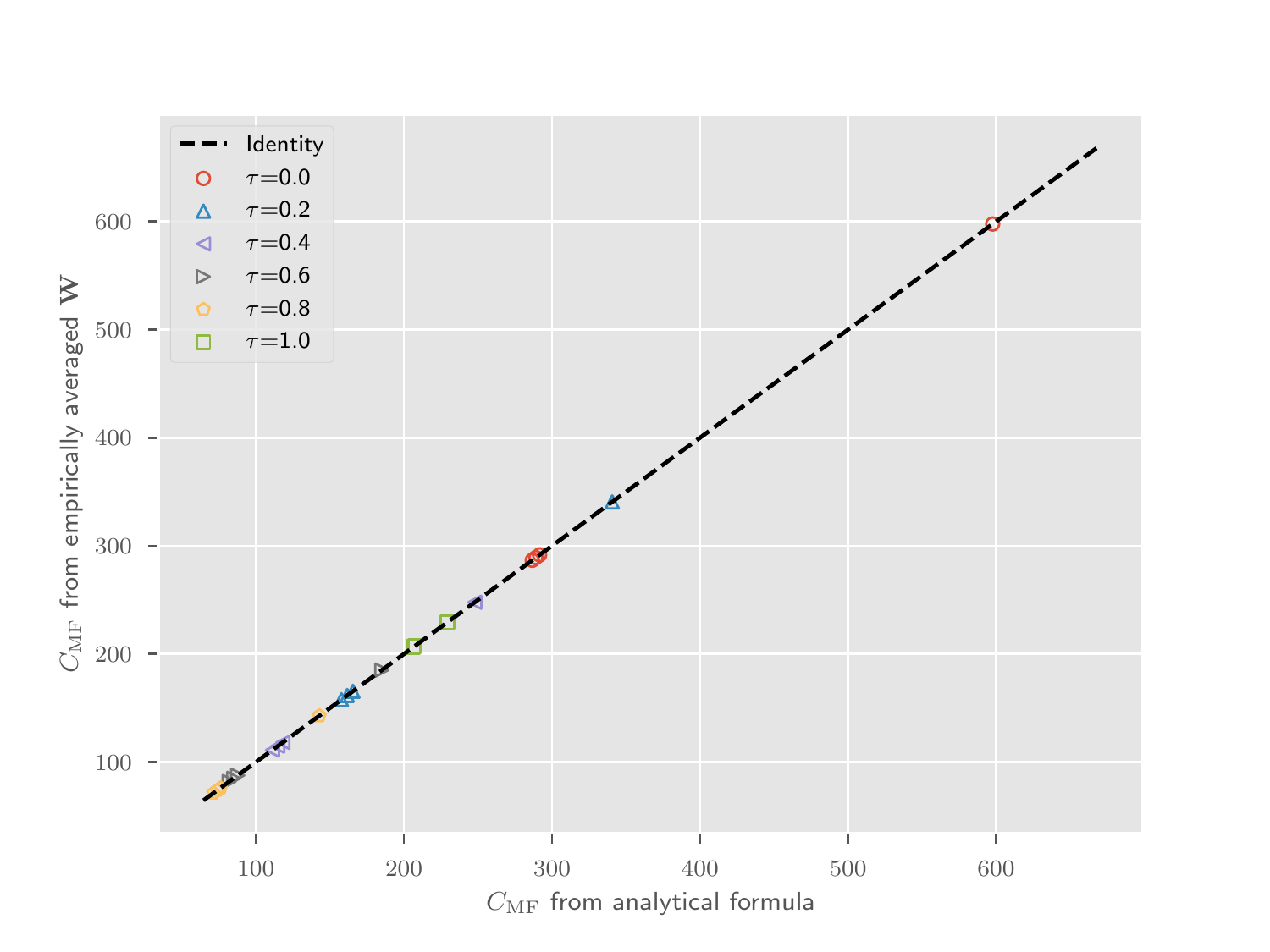}
	\caption{
		Complexity ${\CMF}$ obtained from eq.~\eqref{eq:complexity_full} vs $\CMF$ calculated using eq.~\eqref{eq:mfpt_grounded_laplacian} with an empirically transition matrix $\anglemean{\bW}_\textup{emp}$ averaged over $100$ realisations for each set of parameters. Each datapoint refers to a pair $(\Delta,\tau)$, grouped by colour and symbol according to $\tau$. For all points, we fixed $a=0.7$ and $\Gammap=0.3$.
	}
	\label{fig:MF_MFPT_emp_vs_ana}
\end{figure}

We shall now verify our expression for ${\CMF}$ by comparing it to $\CMF$ obtained from numerical simulations. In these simulations, we fixed a tree with $c=3$ and $h=4$, as well as the parameters $L=48$, $a=0.7$, $\betal=0.07$ and $\Gammap=0.3$. For each given pair of values for $\Delta$ and $\tau$, we calculate $\CMF$ with the following procedure: Generate a 
set of patterns, and subsequently a transition matrix $\bW$ as described in sec.~\ref{sec:model}; repeat 100 times and take the average of the resulting matrices, denoted $\anglemean{\bW}_\textup{emp}$; this average approximates $\anglemean{\bW}$.
The value of $\mrt(\anglemean{\bW}_\textup{emp})$ calculated numerically using eq.~\eqref{eq:mfpt_grounded_laplacian} is our benchmark for $\CMF$ in eq.~\eqref{eq:complexity_full}. The two values of $\CMF$ are plotted against each other in fig.~\ref{fig:MF_MFPT_emp_vs_ana}. We see from the figure that the agreement is excellent, in spite of the fact that the derivation in app.~\ref{app:local_bias_mean_field} uses two approximations (eqs.~\eqref{eq:dv_binomial_approx} and \eqref{eq:bias_expectation_approximation}) to obtain explicit expressions for the entries of $\anglemean{\bW}$.

In the next section, we proceed by testing the goodness of the approximation eq.~\eqref{eq:complexity_MF_def} numerically.

\section{Simulations and Observations\label{sec:simulations}}

This section contains numerical validations of the approximation in eq.~\eqref{eq:complexity_MF_def} by comparing $\CMF$ to $C$, computed numerically as an average over quenched MFPTs $\mrt(\bW)$.
We then proceed to consider the behaviour of $C$ and $\CMF$ as we vary the model parameters. 

For given values of the parameters $a,\Delta$ and $\tau$ we calculate $C(a,\tau,\Delta)$ by repeating the following steps 500 times: Generate a set of patterns, and subsequently the transition matrix $\bW$ according to eq.~\eqref{eq:edge_weight}; record the value $\mrt(\bW)$ calculated numerically using eq.~\eqref{eq:mfpt_grounded_laplacian}. The average of these values is approximately (due to the finite size of the sample) equal to $C(a,\tau,\Delta)$. In these simulations, we fixe a tree with $c=3$ and $h=4$, as well as the parameters $L=48$, $\betal=0.07$ and $\Gammap=0.3$.
With these values for $c$ and $h$, the total number of nodes is $N=121$ and the MFPT for the diffusive random walker (i.e. all edges unweighted) is $\mrt^\textup{diff}=848$.

\begin{figure}[h]
    \centering
    \includegraphics[width=\textwidth]{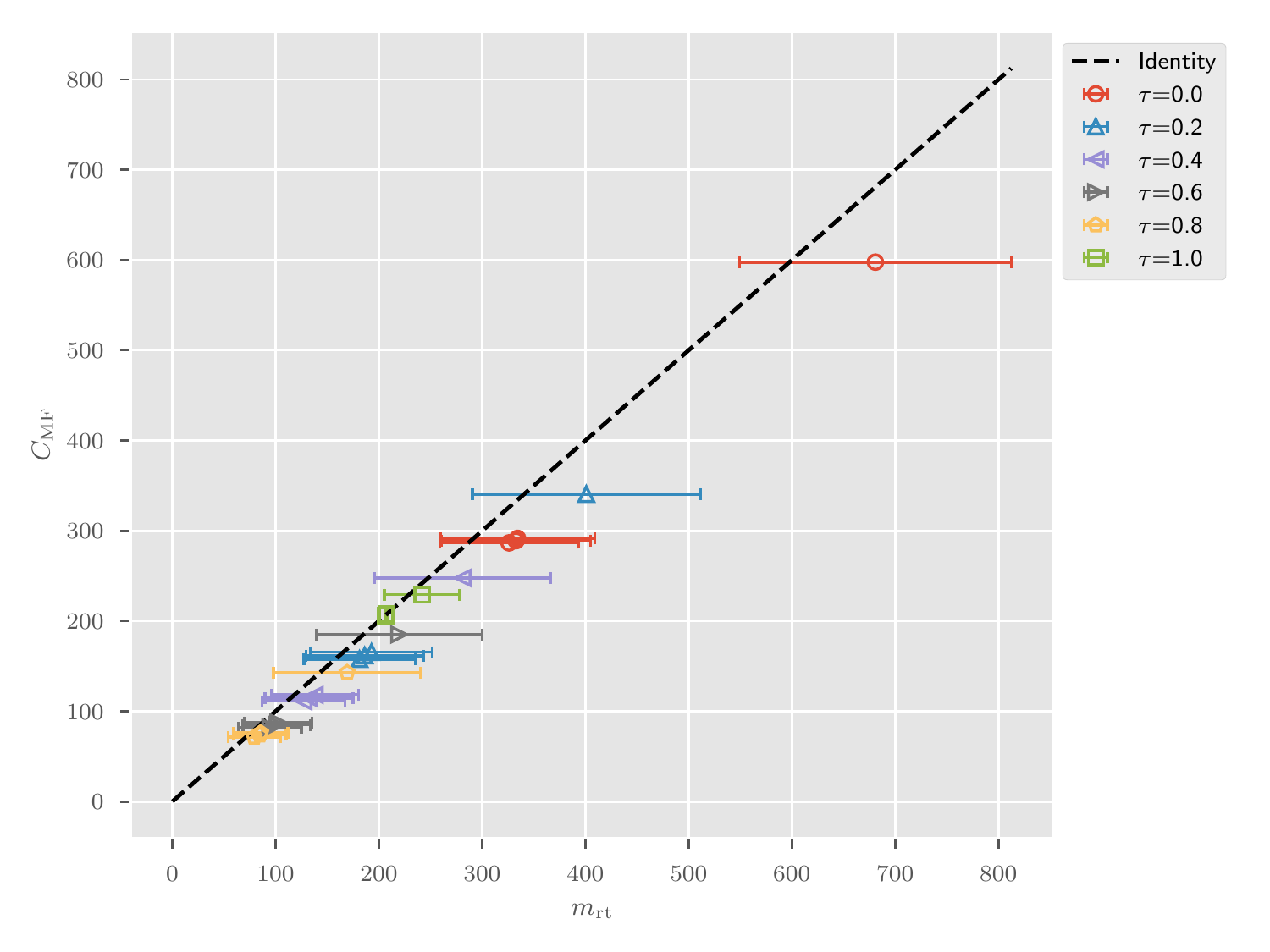}
    \caption{Complexity ${\CMF}$ obtained via sec.~\ref{sec:compfunc} vs $C$, being the average MFPT $\anglemean{\mrt(\bW)}$, where the average is taken over the distribution of patterns. For each realisation of $\bW$, $\mrt(\bW)$ was computed using eq.~\eqref{eq:mfpt_grounded_laplacian}.  Error-bars represent one standard deviation of $\mrt(\bW)$. Each datapoint refers to a pair $(\Delta,\tau)$, grouped by colour and symbol according to $\tau$. For all points, we fixed $a=0.7$ and $\Gammap=0.3$.}
    \label{fig:complexity_scatter}
\end{figure}
In fig.~\ref{fig:complexity_scatter}, we directly compare ${\CMF}$ as per eq.~\eqref{eq:complexity_full} to $C$ in a scatterplot for fixed $a$. The standard deviation of $\mrt(\bW)$ with respect to variations in $\bW$ is indicated as error-bars. The plot confirms for all parameters considered that eq.~\eqref{eq:complexity_MF_def} leads to a systematic underestimation while accurately reflecting the correct trend.

\begin{figure}[h]
    \centering
    \includegraphics{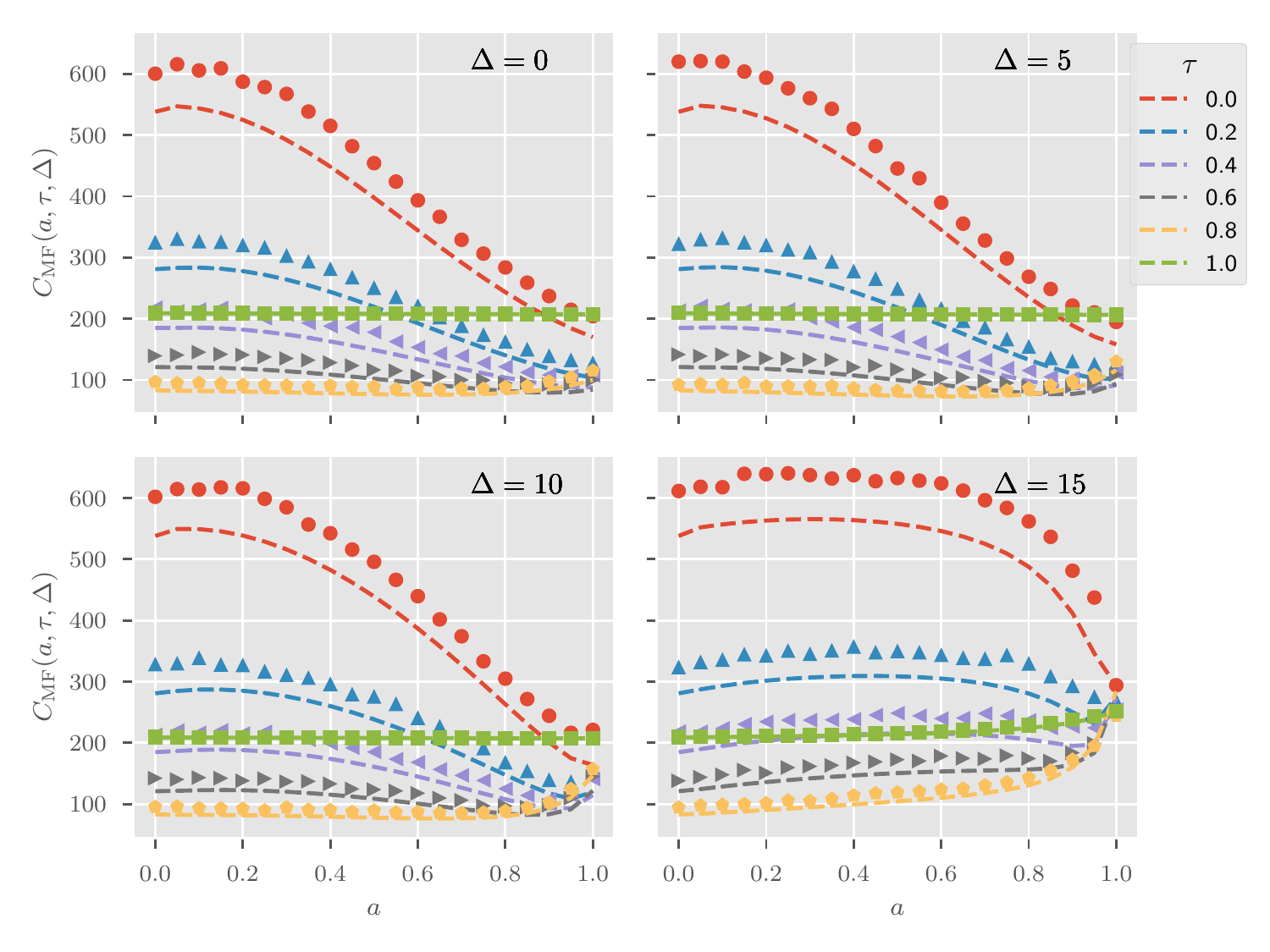}
    \caption{Complexity ${\CMF}$ as a function of $a$ for different values of $\tau$ and $\Delta$. The points show the numerical average for $C$ on the right hand side of eq.~\eqref{eq:complexity_MF_def}, taken over $500$ realisations of $\bW$.}
    \label{fig:complexity_a_tau_Delta}
\end{figure}
Next, we analyse the dependence of ${\CMF}$ on the different parameters of our model. 
Fig.~\ref{fig:complexity_a_tau_Delta} plots ${\CMF}$ and $C$ as a function of $a=\anglemean{\xi_j}$, the expectation of root-level bits $\xi_j$. The dashed lines represent the mean-field approximation $\CMF$ in eq.~\eqref{eq:complexity_full}, the symbols the value of $C$ as obtained in the beginning of this section. 
Fig.~\ref{fig:complexity_a_tau_Delta} confirms that ${\CMF}$ tracks $C$ faithfully, also for varying $a$. 

For small values of $\tau$, i.e. little vertical coherence between patterns, the complexity ${\CMF}$ first shows a slight increase as a function of $a$ (approximately in the interval $0<a<0.2$) before a more pronounced decrease to about $1/3$ of its maximum (for $a>0.2$). As $\tau$ approaches $1$, the curves for ${\CMF}$ first become notably flatter and lower, which is as expected since higher vertical coherence is more likely to put the reader on the right track faster. This effect is only seen up to $\tau=0.8$, as too high coherence -- given if $\tau\to 1$ -- forces all patterns within the same Part to be equal, which does not help the reader navigate at all. 

Fig.~\ref{fig:complexity_a_tau_Delta} suggest the following conclusion: For fixed, low values of $\tau$, the complexity can be minimised by increasing $a$ as much as possible. Since $a$ represents the keyword density of the root pattern, this means that the Title of the represented Act should reference as many keywords as possible. For high values of $\tau$, the complexity increases with $a$, though the increase is far less pronounced than the decrease at low $\tau$.

\begin{figure}[h] 
    \centering
    \includegraphics{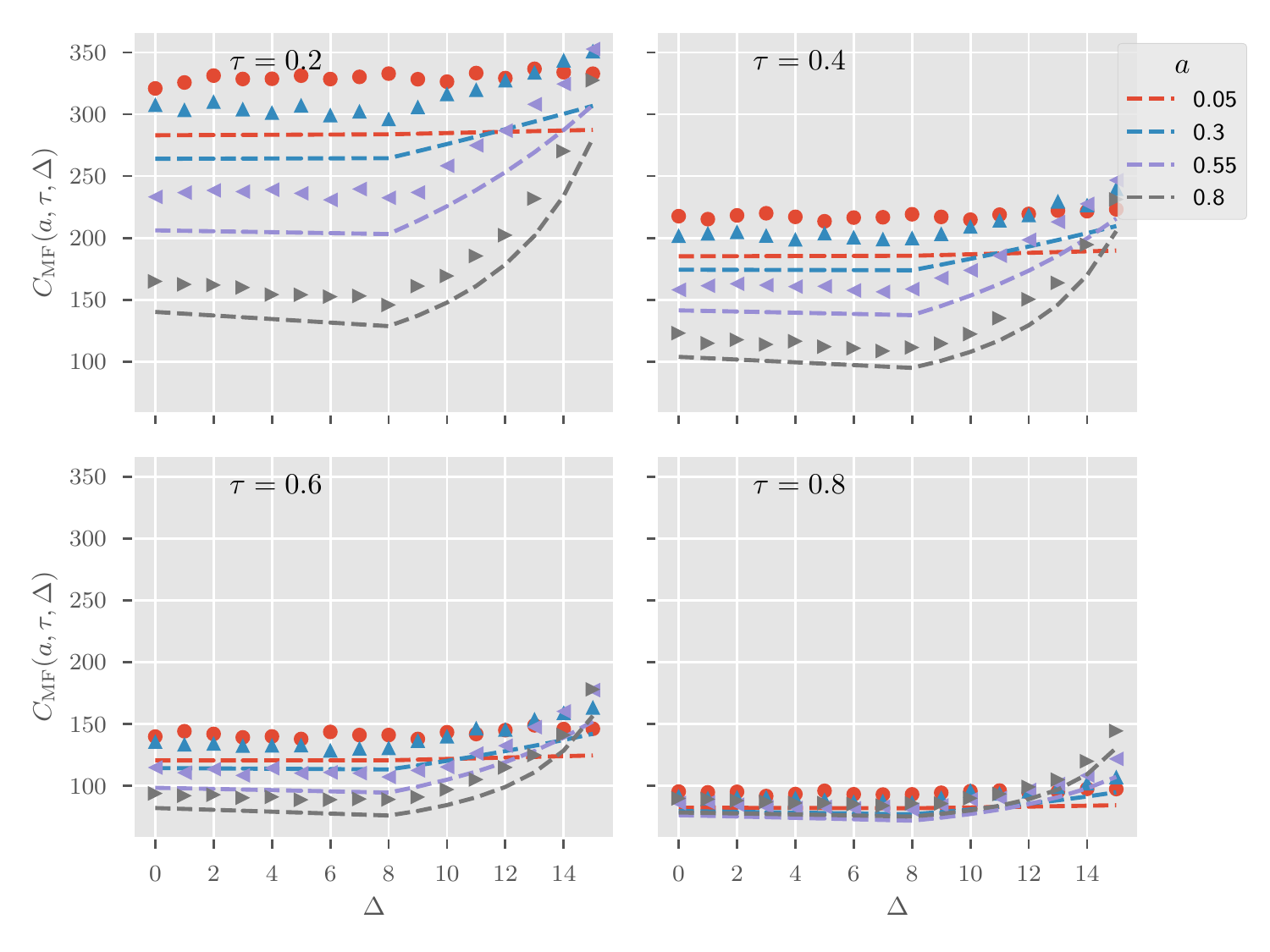}
    \caption{Complexity ${\CMF}$ as a function of $\Delta$ for different values of $a$ and $\tau$. The points show the numerical average for $C$ on the right hand side of eq.~\eqref{eq:complexity_MF_def}, taken over $500$ realisations of $\bW$.}
    \label{fig:complexity_Delta_a_tau}
\end{figure}
Fig.~\ref{fig:complexity_Delta_a_tau} shows ${\CMF}$ as dashed lines and $C$ as symbols as functions of $\Delta$. The panels and different curves per panel correspond to different values for $a$ and $\tau$, respectively. We make the same observation as above about the systematic underestimation incurred in eq.~\eqref{eq:complexity_MF_def}, although in addition to $\tau$, the offset also seems to decrease with $a\to 1$. 

${\CMF}$ is largely constant in $\Delta$ for $\Delta\leq 8$.
Beyond this value, $C$ begins to increase with $\Delta$, except for the lowest tested value $a=0.05$. $C$ increases by about a factor of $2$ for $\Delta>8$. Strikingly, ${\CMF}$ and $C$ show a slight \emph{decrease} up to $\Delta\leq 8$ for $a=0.8$, which is contrary to our intuition that higher overlap between adjacent Parts should increase 
the complexity
as it leads to more initial missteps of the random walker. However, the observed decrease is minor compared to the observed increase exhibited at higher $\Delta$. The fact that such increase in complexity is less significant for higher $\tau$ is again in line with our expectation that a more ``vertically coherent'' text should be overall easier to navigate. We note that the range of values of $C$ over $\Delta$ is less than the one over $a$, shown in fig~.\ref{fig:complexity_a_tau_Delta}, showing that $\Delta$ has lower influence. 

We deduce from fig.~\ref{fig:complexity_Delta_a_tau} that in order to reduce $C$, $\Delta$ should not be chosen too high. Further simulations suggest that depending on the coordination number $c$ of the tree, the complexity can also rise if $\Delta$ is chosen too low. This means that $C$ has a local minimum in $\Delta$, which represents the optimal keyword overlap between adjacent Parts. 

\begin{figure}[h]
    \centering
    \includegraphics{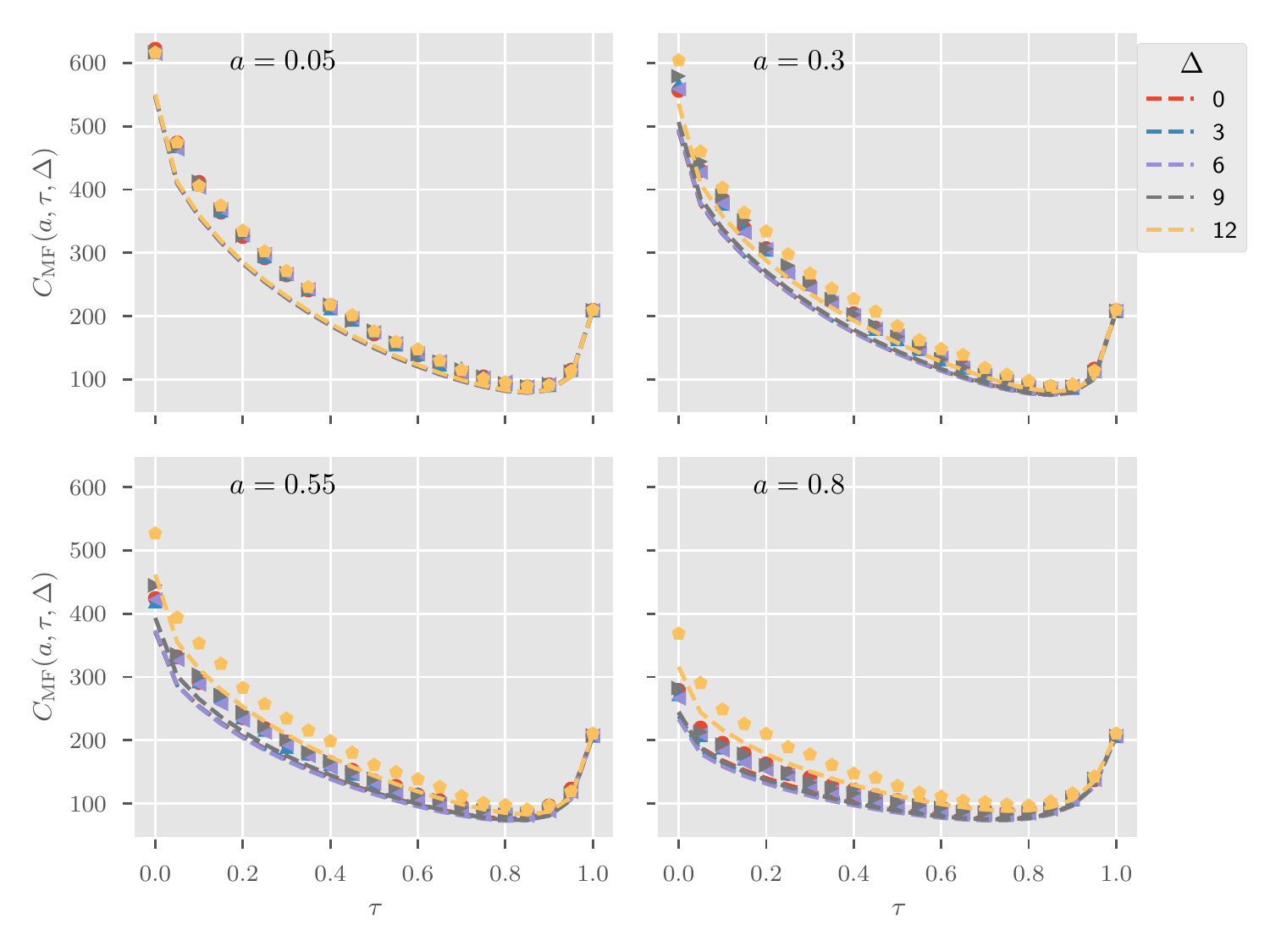}
    \caption{Complexity ${\CMF}$ as a function of $\tau$ for different values of $\Delta$ and $a$. The points show the numerical average for $C$ on the right hand side of eq.~\eqref{eq:complexity_MF_def}, taken over $500$ realisations of $\bW$.}
    \label{fig:complexity_tau_Delta_a}
\end{figure}
Fig.~\ref{fig:complexity_tau_Delta_a} presents ${\CMF}$ as a function of $\tau$ as dashed lines, with different curves and panels represent different values for $\Delta$ and $a$, respectively. Different symbols are used to represent the values of $C=\anglemean{\mrt(\bW)}$. 

The figure shows that ${\CMF}$ has a pronounced local minimum in $\tau$ between $0.8$ and $0.9$ for all values of $\Delta$ and $a$ tested.
At $\tau=1$, the random walker becomes diffusive whenever inside a Part, because all patterns within a given Part are identical. Therefore it makes sense that the minimal complexity should not be realised at this value, since the patterns cease to guide the walker to $\targetnode$. The range of variation of $C$ over $\tau$ is comparable to that over $a$, as shown in fig.~\ref{fig:complexity_a_tau_Delta}.

To conclude the analysis of fig.~\ref{fig:complexity_tau_Delta_a} we summarise that $C$ may be minimised by choosing an appropriately high value for $\tau$, which should not be too close to $1$. This means that one should allow the keywords within one Part to vary slightly, to avoid keyword patterns that are either almost identical or approximately independent.

Figs.~\ref{fig:complexity_a_tau_Delta}, \ref{fig:complexity_Delta_a_tau} and \ref{fig:complexity_tau_Delta_a} suggest that
an increasing $\Delta$, or decreasing $\tau$ or $a$ results in
a higher rate of mistakes made by the walker, thus leading to a higher searching time. 
Since $\Delta$ has its primary effect on the root (Act) level, it is dominated by $\tau$, which controls noise on all (bar the Act) levels. Since $a$ also affects the values of $a_h$ and $a_l$, it has an effect on all levels as well; accordingly, we observe that varying $a$ and $\tau$ have comparable effects on $C$, and dominate variations over $\Delta$.
Form here we conclude that the priority should be on maximising the tightness $\tau$ and keyword frequency $a$ to reduce the complexity of the modelled ensemble of Acts.

This section shows that $\CMF$ faithfully reproduces the trends of $C$ for varying $\tau$, $\Delta$ and $a$. This entails a significant benefit because it allows us to optimise the parameters of the model with respect to $\CMF$ without the need for costly simulations. As a consequence, one can imagine optimising a real legal text, by estimating its parameter values for our model and tweaking the text and layout to minimise $\CMF$. 

Here we have not considered the effects of the parameters $c$, $h$, $L$ and $\Gammap$. The former two of these deserve a word of caution: The number of nodes increases as $c^{h+1}$, and $\mrt$ for the regular random walker as $hc^{h+1}$. Therefore, without appropriate rescaling, the values of $C$ and ${\CMF}$ do not allow for the comparison of graphs of different size.

\section{Conclusions and Outlook\label{sec:conclusio}}

We presented a quantitative theory of informational complexity of legal trees by analysing a random walker model for the retrieval of information planted in the leaves of a legal tree. The model assumes that the reader proceeds by keyword affinity, such that it is drawn towards nodes whose content looks similar to the target information. The searched text is generated randomly, with two main parameters controlling its horizontal and vertical coherence. Our analysis and numerical simulations show that these properties of the text have the desired effect on the random walker: With high vertical coherence, the content of the leaves is well-reflected in the top items (Parts) of the text, and the reader finds its target more quickly. High horizontal coherence, on the other hand, means that different Parts are difficult to discern, leading to more initial errors by the reader.

As a measure of complexity, we propose the MFPT of the random reader from the root of the tree to the predefined target information; it gives an intuitive account of how difficult it is for a typical reader to navigate the legal text by following only local information. Similarly, MFPTs have also been employed successfully to asses the heterogeneity and transport properties of social and other complex networks \cite{Bassolas2021}.

So far, we have limited our analysis to trees, where we were able to compute our complexity measure analytically using simple approximations. However, other topologies play an important role in real-life legal networks. A direct generalisation of the present work can be the inclusion of cross-references, which can potentially lead to detours \emph{and} act as shortcuts. 
In fact, studies of European civil law have found that these legal systems can exhibit small-world properties \cite{Koniaris2018}. 
In other studies, the more general directed acyclic graphs are used to represent citation networks, e.g. the network of precedents in the US \cite{Bommarito2010} or of the total citation network in a system of statute law \cite{Bommarito2011,Katz2020}. Recently, \cite{Coupette2021} has elaborated on the similarity between legal and software systems, drawing from best practices on the latter to propose improvements on the former. The framework developed in the present paper may prove useful in providing a more quantitative ground to assess the methods in these lines of research as well.

Moreover, our model definitions rely on a number of assumptions on the details of how keywords are distributed  over the text: Firstly, the definition of overlap assumes that the Parts of an Act be ordered in such a way that consecutive pairs realise the maximal overlap in that Act, and that this overlap is the same for all adjacent pairs. Secondly, we assume that below the Part level, the marginal distributions for every keyword is fixed within each part, which might be unrealistic for ``deep'' laws with many levels. Relaxing these assumptions -- introduced for the sake of computational simplicity -- may render the model even more realistic and general.

We have modelled a reader as a Markovian random walker, that is to say that it is ``memoryless''. To replicate the behaviour of a real reader more closely, more general types of walks (e.g. self-avoiding walks \cite{LopezMillan2012}) might be appropriate. 

Finally, on the side of our analysis, it should be possible to refine the approximation in eq.~\eqref{eq:complexity_MF_def}. Perhaps, more of the information contained in the pattern-dependent $\mrt(\bW)$ can be exploited by carrying the analysis beyond its mean to higher moments. 

Previous research indicates that glossaries (lists of keywords) may be extracted using natural language processing \cite{Mihalcea2011} (in particular topic models \cite{Carlson2020}). To make our model applicable in real-life scenarios, one should devise a way to estimate the horizontal and vertical coherence of an existing legal document with a tree-like backbone, after a glossary has been extracted. 

We can derive three broad and intuitive lessons from the results in sec.~\ref{sec:simulations} to reduce the ``complexity'' of a legal tree. (i) Keywords at the lowest levels should be reflected at higher levels, i.e. a legal text should be ``tightly'' formulated. Yet, it is possible to make it overly tight, which happens when all text items look too similar to each other. This situation is identical to giving no clues at all to the reader. (ii) Parts should be well separated by their keywords (and hence by topic); some keyword overlap is acceptable, as long as sufficiently many Part-specific keywords remain to guide the reader. (iii) Text at higher levels should not be too sparse. If high-level entries contain only a small number of keywords (such as a short headline), little information about its subordinate items can be conveyed (except by interpretation, e.g. through association of keywords and related words). A higher keyword frequency at the top levels saves the reader time-costly detours into wrongs Parts.

\begin{acknowledgments}
PV and ET acknowledge support from UKRI Future Leaders Fellowship scheme [n. MR/S03174X/1]. Y-PF is supported by the EPSRC Centre for Doctoral Training in Cross-disciplinary Approaches to Non-Equilibrium Systems (CANES EP/L015854/1).
The authors acknowledge use of the research computing facility at King’s College London, \textit{Rosalind} (\url{https://rosalind.kcl.ac.uk}).
\end{acknowledgments}


\appendix

\section{Expected Pattern Distances}\label{app:expected_pattern_distances}

In this appendix we derive the expected pattern distances eqs.~\eqref{eq:overlap_expected_distance} and \eqref{eq:tightness_expected_distance} in sec.~\ref{sec:expected_pattern_distances}. Before we start, it will be useful to recall that $X$ is a binomial random variable (denoted $X\sim \binomrv{p}{n}$) if it has the probability mass function (PMF)
\begin{equation}\label{eq:binomial_pmf}
    \prob{X=x}=\binom{n}{x}p^{x}(1-p)^{n-x}
\end{equation}
with expectation
\begin{equation}\label{eq:binomial_expectation}
    \mean(X)=np\ .
\end{equation}

We begin with the expectation of
the Hamming distance on Part-level, $d^{\mu,\mu+1}:=d\left( \bm\xi^\mu,\bm\xi^{\mu+1} \right)$. From the model definitions in sec.~\ref{sec:model}, we see that this is a combination of binomial random variables, for either $\xi^\mu_j = \xi^{\mu+1}_j$ or $\xi^\mu_j \neq \xi^{\mu+1}_j$, and the probabilities of both events depend on whether both bits have the same marginal expectation $\ah$ (or $\al$), or if one is equal to $\ah$ and the other equal to $\al$.

Let the numbers of indices $j$ 
such that $\anglemean{\xi^\mu_j}=\anglemean{\xi^{\mu+1}_j}=a_h$ and $\anglemean{\xi^\mu_j}=\anglemean{\xi^{\mu+1}_j}=a_l$
be $L_{hh}$ and $L_{ll}$, respectively, and let the number of indices with $\anglemean{\xi^\mu_j}\neq\anglemean{\xi^{\mu+1}_j}$ be $L_{hl}$. The $L$'s are functions of $\Delta$ (cf. fig.~\ref{fig:overlap_schema})
\begin{align}
    \begin{array}{lll}
        L_{hh}(\Delta) &= 
             \left\{ 
             \begin{array}{l}
                2\Delta \\
                4\Delta-\lc (c-2) 
             \end{array}
             \right. 
             & 
             \begin{array}{l}
                  \colon \Delta \leq \lc \frac{(c-2)}{2}\ , \\
                  \colon \ \textup{else}\ ,
             \end{array}
              \\
        L_{ll}(\Delta) &=
            \left\{
            \begin{array}{l}
                \lc(c-2) -2\Delta \\
                0
            \end{array}
            \right.
            &
            \begin{array}{l}
                 \colon \Delta \leq \lc \frac{(c-2)}{2}\ , \\
                 \colon \ \textup{else}\ , 
            \end{array}
            \\
        L_{hl}(\Delta) &=
            \left\{
                \begin{array}{l}
                     2\lc \\
                     2\lc(c-1)-4\Delta 
                \end{array}
            \right.
            &
                \begin{array}{l}
                     \colon \Delta \leq \lc \frac{(c-2)}{2}\ , \\
                     \colon \ \textup{else}\ ,
                \end{array}
    \end{array}
\end{align}
where $L$ is the number of bits per patterns, $c$ is the number of Parts, and $\lc=L/c$.
The reason for the presence of different cases is that the inequality
\begin{equation}
    (\mu-1)\lc-\Delta+L < (\mu+1)\lc+\Delta 
\end{equation}
is true if and only of $\Delta>\lc \frac{c-2}{2}$. Now $(\mu-1)\lc-\Delta+L$ is the ``left'' boundary of the $\ah$-domain of Part $\mu$ after applying the $L$-periodicity of eq.~\eqref{eq:ah_al_def}, and $(\mu+1)\lc+\Delta$ is the ``right'' boundary of the $\ah$-domain of Part $\mu+1$; therefore, $\Delta\leq\lc\frac{c-2}{2}$ implies $\anglemean{\xi^\mu_j}=\anglemean{\xi^{\mu+1}_j}=\ah$
if and only if $\mu\lc-\Delta < j \leq \mu\lc+\Delta$,
while for $\Delta > \lc \frac{c-2}{2}$ there is a second set of solutions given by $(\mu-1)\lc-\Delta+L < j \leq (\mu+1)\lc + \Delta$ (refer to fig.~\ref{fig:overlap_schema} for a schematic illustration). Notice that with $\Delta>\lc\frac{c-2}{2}$, we necessarily have $L_{ll}=0$, i.e. the $\ah$-domains of both patterns together cover all of $\{1,\dots,L\}$.

The probabilities of $\{\xi^\mu_j \neq \xi^{\mu+1}_j\}$ can be derived using the law of total probability
and the conditional independence of the Part patterns given the root pattern $\bxi$, 
\begin{align}
 \nonumber   
    \prob{\xi^\mu_j \neq \xi^{\mu+1}_j}=& \sum_{z=0}^1 \prob{\xi^\mu_j \neq \xi^{\mu+1}_j \mid \xi_j=z }\prob{\xi_j=z}
 \\
 =&\sum_{z=0}^1 \prob{\xi_j=z}\Big[\prob{\xi^\mu_j=z \mid \xi_j=z } \prob{ \xi^{\mu+1}_j \neq z \mid \xi_j=z }
 \nonumber\\ 
    &+  \prob{\xi^\mu_j \neq z \mid \xi_j = z } \prob{ \xi^{\mu+1}_j = z \mid \xi_j=z }\Big]
    \label{eq:parts_neq_total_prob}
    \ ,
\end{align}
where the factors in the square brackets are given by the elements of the $\bR_j$'s in eq.~\eqref{eq:R_def} for each of the combinations of $\ah$ and $\al$.
In fact, by definition of $\bR_j$, we have
\begin{align}
    \nonumber
    \prob{\xi^\mu_j \neq \xi^{\mu+1}_j \mid \xi_j=0 }=&\bR^{\mu}_j(0\vert 0) \bR^{\mu+1}_j(1\vert 0)+\bR^{\mu}_j(1 \vert 0) \bR^{\mu+1}_j(0 \vert 0) \\ \nonumber
    =&2(1-\Gammap)\Gammap\\
    \prob{\xi^\mu_j \neq \xi^{\mu+1}_j \mid \xi_j=1 }=&\bR^{\mu}_j(0 \vert 1)\bR^{\mu+1}_j(1 \vert 1)+\bR^{\mu}_j(1 \vert 1)\bR^{\mu+1}_j(0 \vert 1) \\ \nonumber
    =& \frac{a-a^\mu_j+(1-a)\Gammap}{a}\frac{a^{\mu+1}_j-(1-a)\Gammap}{a}\\ 
    &+\frac{a^\mu_j-(1-a)\Gammap}{a}\frac{a-a^{\mu+1}_j+(1-a)\Gammap}{a} \ .
\end{align}
With the decomposition eq.~\eqref{eq:parts_neq_total_prob}, and using that $\anglemean{\xi_j}=\prob{\xi_j=1}=a$ for all $j$, this produces the marginal probability
\begin{align}
    \nonumber
    \prob{\xi^\mu_j \neq \xi^{\mu+1}_j}=&2(1-a)\Gammap(1-\Gammap)\\ \nonumber
    &+\frac{1}{a}\left[
    (a-a^\mu_j+(1-a)\Gammap )(a^{\mu+1}_j-(1-a)\Gammap)\right.\\ 
    &+\left.
    (a^\mu_j-(1-a)\Gammap)(a-a^{\mu+1}_j+(1-a)\Gammap)
    \right]\ .
\end{align}
Considering now all $L_{hh}$ bits $j$ for which $\anglemean{\xi^\mu_j}=\anglemean{\xi^{\mu+1}_j}=a_h$, the above expression
reduces to $2(1-a)(1-\Gammap)\Gammap$ since $a_h=(1-a)\Gammap+a$ (given in eq.~\eqref{eq:ah_betah}) sets the other two summands to zero. The sum of the distances $\vert \xi^\mu_j-\xi^{\mu+1}_j \vert$ for theses bits then forms a binomial random variable with ``success'' probability $2(1-a)(1-\Gammap)\Gammap$. Similarly, we can treat the other bits in two groups of size $L_{lh}$ and $L_{ll}$, respectively, as described in the beginning of this section. For this purpose, it is useful to recall from eq.~\eqref{eq:al_betal} that $a_l-(1-a)\Gammap=\beta_l$.
The total distance $d^{\mu,\mu+1}$ is then given by a sum of binomial random variables
\begin{align}
    \nonumber    
    d^{\mu,\mu+1} \sim& \binomrv{ 2(1-a)\Gamma^\prime(1-\Gamma^\prime)}{ L_{hh} } \\ \nonumber
        &+ \binomrv{ 2(1-a)\Gamma^\prime(1-\Gamma^\prime) +a-\betal  }{ L_{hl} } \\ 
        &+ \binomrv{ 2(1-a)\Gamma^\prime(1-\Gamma^\prime) +2\frac{\left(a-\betal\right)\betal}{a} }{ L_{ll} }\ ,
\end{align}
expressed in terms of $\beta_l$ for conciseness. 

The expected pattern-distance between neighbours is readily calculated using the linearity of $\anglemean{\cdot}$, the above characterisation for $d^{\mu, \mu+1}$ and the expectation for 
binomial random variables eq.~\eqref{eq:binomial_expectation}. These ingredients give the result reported for $\bad=\anglemean{d^{\mu,\mu+1}}$ in eq.~\eqref{eq:overlap_expected_distance}.
We can see that $\bad$ is a decreasing function in $\Delta$ with maximum and minimum 
\begin{align}
    \badmax =& \bad_0 + 2\lc (a-\betal)\left( (c-2)\frac{\betal}{a} + 1\right)\ , \\
    \badmin =& \bad_0\ ,
\end{align}
with $\bad_0=2(1-a)\Gammap(1-\Gammap)L$.
Moreover, $\bad$ is a combination of two affine linear function, with the transition between the two occuring at $\Deltatrans=\lc\frac{c-2}{2}$ with value
\begin{align}
    \badtrans= \bad(\Deltatrans)= \bad_0 + 2\lc (a-\betal)\ .
\end{align}

We now examine the expected distance of two patterns $\bxi^\mu$, $\bzeta$, where the former 
represents
Part $\mu$, and the 
latter is a
descendant of the former, at distance $k$. We are particularly interested in the situation where $\bzeta$ is a leaf-level pattern, i.e. the edge-distance between the two is $k=h-1$. To determine $\anglemean{d(\bzeta, \bxi^\mu)}$, we need to know the probabilities of the events $\{ \xi_j^\mu \neq \zeta_j \}$ which we calculate from the $k$-th power of $\bQ$ defined in eq.~\eqref{eq:Q_def}
\begin{equation}\label{eq:Q_power}
    \bQ^{k}(a_j^\mu) =\begin{pmatrix}
        1-a_j^\mu\left(1-(1-\Gamma)^k\right) & a_j^\mu \left(1-(1-\Gamma)^k\right)\\ (1-a_j^\mu) \left(1-(1-\Gamma)^k\right)
         & 1-(1-a_j^\mu) \left(1-(1-\Gamma)^k\right)
        \end{pmatrix}\ ,
\end{equation}
where $a_j^\mu$ is the marginal expectation 
$\anglemean{\xi_j^\mu}=a_j^\mu$. As $a_j^\mu$ can take the values $\al$ and $\ah$, as defined in eqs.~\eqref{eq:al_betal} and \eqref{eq:ah_betah}, the $j$-th bits are different with probability
\begin{align}\label{eq:tightness_single_bit_difference_prob}
    \nonumber    
    \prob{ \xi_j^\mu \neq \zeta_j } &= \prob{\zeta_j=0 | \xi_j^\mu=1} \prob{\xi_j^\mu=1} + \prob{\zeta_j=1 | \xi_j^\mu=0} \prob{\xi_j^\mu=0} \\\nonumber
        &=\bQ^k ( 0 | 1 ) a_j^\mu+\bQ^k ( 1 | 0 ) (1-a_j^\mu) \\ 
        &= 2a_j^\mu(1-a_j^\mu)\left[ 1-\left( 1-\Gamma \right)^{k} \right]
        =:g^{(k)}(a_j^\mu)\ .
\end{align}

For the full pattern-distance composed by all bits, we have to take both possible values, 
$\al$ and $\ah$, for $a_j^\mu$ into account. Again, the full pattern-distance is a sum of two independent binomial random variables
\begin{equation}
    d(\bm{\zeta},\bxi^\mu) \sim \binomrv{g^{(k)}(\ah)}{ \lc+2\Delta } +  \binomrv{g^{(k)}(\al)}{ \lc (c-1)-2\Delta }\ ,
\end{equation}
with expectation given by eq.~\eqref{eq:binomial_expectation},
\begin{equation}\label{eq:vertical_pattern_distance}
    \anglemean{d(\bm{\zeta}, \bxi^\mu ) } = g^{(k)}(\al) \left(\lc (c-1)-2\Delta\right) + g^{(k)}(\ah) \left(\lc+2\Delta\right)\ .
\end{equation}
For 
$k=h-1$,
we obtain the result given in the main text in eq.~\eqref{eq:tightness_expected_distance}.

\section{$5$-Tuple Labels for Paths and Nodes\label{app:path_labels}}

The statistical properties of the pattern-distance of any node to the target are determined by the location of the node in the tree. The determining criteria are
(i) the edge-distance between the node and the target, (ii) whether the root lies on the shortest path between the node and the target, and (iii) the Part containing the node, as 
Part patterns 
have different switching probabilities over the edges incident to the root. 

We will label directed paths along the edges of the tree by $5$-tuples, which encode features of the paths that influence the statistical properties of pattern-distance between
the start and end nodes of these paths. 

\begin{figure}[H]
    \centering
    \includegraphics{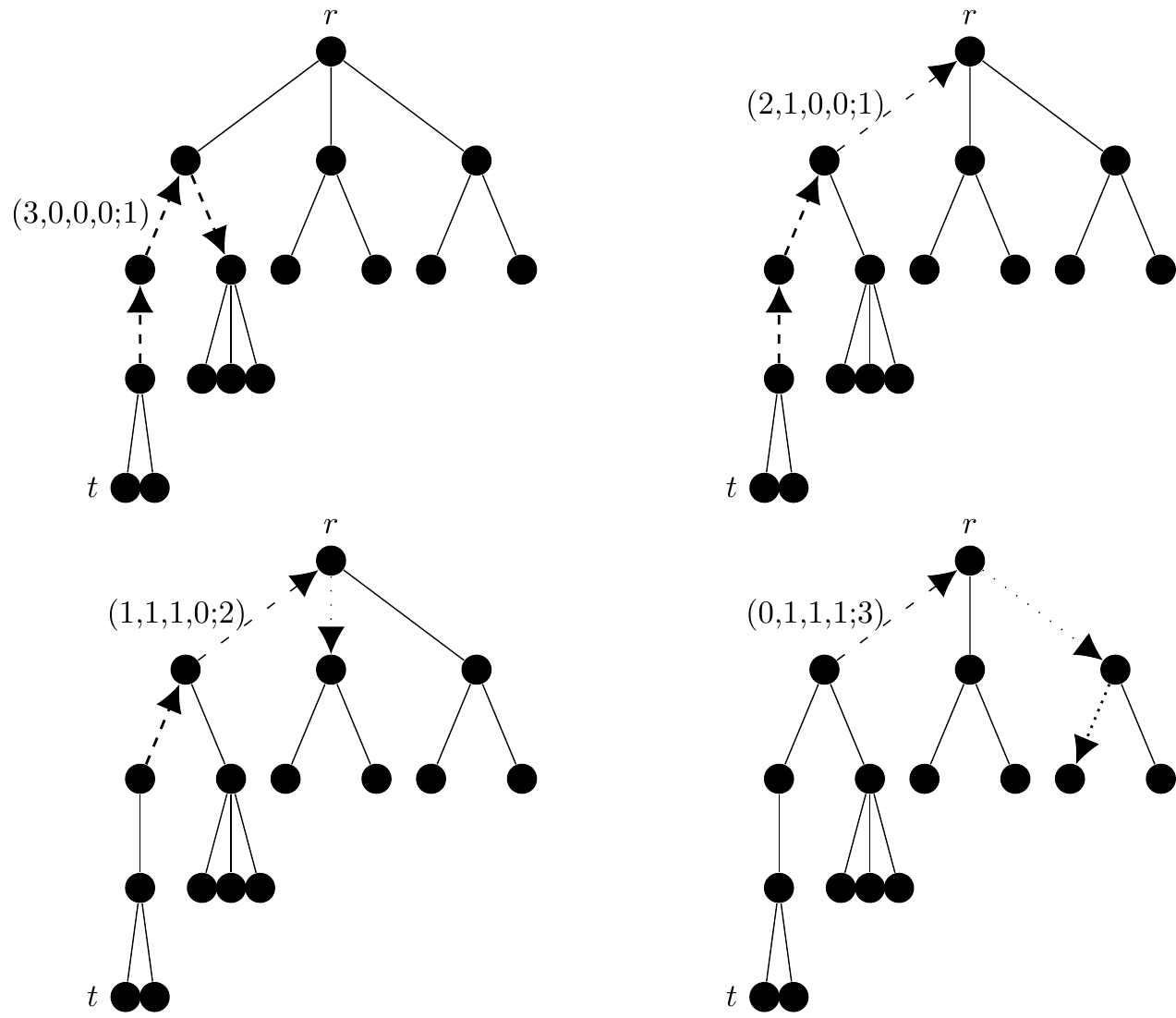} 
    \caption{Labels for different paths of length $3$ oriented away from $\targetnode$, shown by arrows. The number of densely dashed and densely dotted arrows give the first and fourth coordinate, respectively. The number of thinner, loosely dashed and loosely dotted arrows give the second and third coordinates, respectively. The latter two are always associated with the root, so the third and fourth entries are both either $0$ or $1$. 
    Note that the first four coordinates always sum to the path-length.\label{fig:path_labels}}
\end{figure}

Let $P$ a directed path of length $p$ such that
at no point following its direction, one moves closer (in the sense of edge-distance) to $\targetnode$ (see fig.~\ref{fig:path_labels} for examples).
Without loss of generality, we may assume that the Parts ${\mu}$ are enumerated such that $1$ is the Part containing $\targetnode$. The remaining Parts may be in any order. We assign to $P$ a $5$-tuple $P\sim (P_1,P_2,P_3,P_4;P_5)$, where 
$P_1$ is the number of edges of $P$ connecting its starting node to the Part node $\mu=1$. We set $P_2=1$ if the edge $(1,\rootnode)$ from Part 1 to the root lies on $P$, and otherwise we have $P_2=0$. We define $P_3=1$ if $P$ has an edge $(\rootnode,\mu)$ from the root to a Part node with $\mu\geq 2$, and otherwise we set $P_3=0$. $P_4$ counts the number of edges of $P$ connecting its end node to the nearest Part node $\mu \geq 2$. $P_5$ records the Part $\mu$ containing the end node of $P$.

In practice, the label $(P_1,P_2,P_3,P_4;P_5)$ can be determined by making the following observation.
If the root node $\rootnode$ lies on $P$ and $\ell$ is the length of the path $P$,  
there are two numbers $k,m\geq 0$ such that $P$ consists of $k$ edges in Part $1$ and $m=\ell-k$ edges in exactly one other Part, say $\mu$ (we count the edge $(\rootnode,\mu)$ as belonging to Part $\mu$). By construction, we have $k=0$ if and only if $P$ starts in the root node $\rootnode$ and $m=0$ if and only if $P$ ends in $\rootnode$.
Then, we label $P$ as
\begin{align}
    P \sim \begin{cases}
        (k-1,1,0,0;1) \quad & \colon P \textup{ ends with } \rootnode, \\
        (0,0,1,m-1;\mu) \quad & \colon P \textup{ begins with } \rootnode, \\
        (k-1,1,1,m-1;\mu) \quad & \colon \textup{else}.
    \end{cases}
\end{align}
If $\rootnode$ does not lie on $P$, we assign the label
\begin{align}
    P \sim \begin{cases}
        (p,0,0,0;1) \quad & \colon P \textup{ lies in the target-Part } \ , \\
        (0,0,0,p;\mu) \quad & \colon P \textup{ lies in Part $\mu>2$.}
    \end{cases}
\end{align}
Various examples for labelled paths are given in fig.~\ref{fig:path_labels}.
In all cases, the sum of the first four indices equals the length of the path $P$. 

We anticipate that our notation will not be well defined for most directed paths in the tree, but it will be well defined for those paths relevant to the analysis in this manuscript.
Moreover, the above definition
does not identify paths uniquely; for instance, the label $(k,0,0,0;1)$ applies to all paths of length $k$ not leaving the target-Part and with the additional constraint of being directed away from $\targetnode$. However, due to the constraint on the path direction, the pattern-distances between the start and end nodes of each path are identically distributed, and the distribution is determined by $(k,0,0,0;1)$.

\begin{figure}[h]
    \centering
    \includegraphics{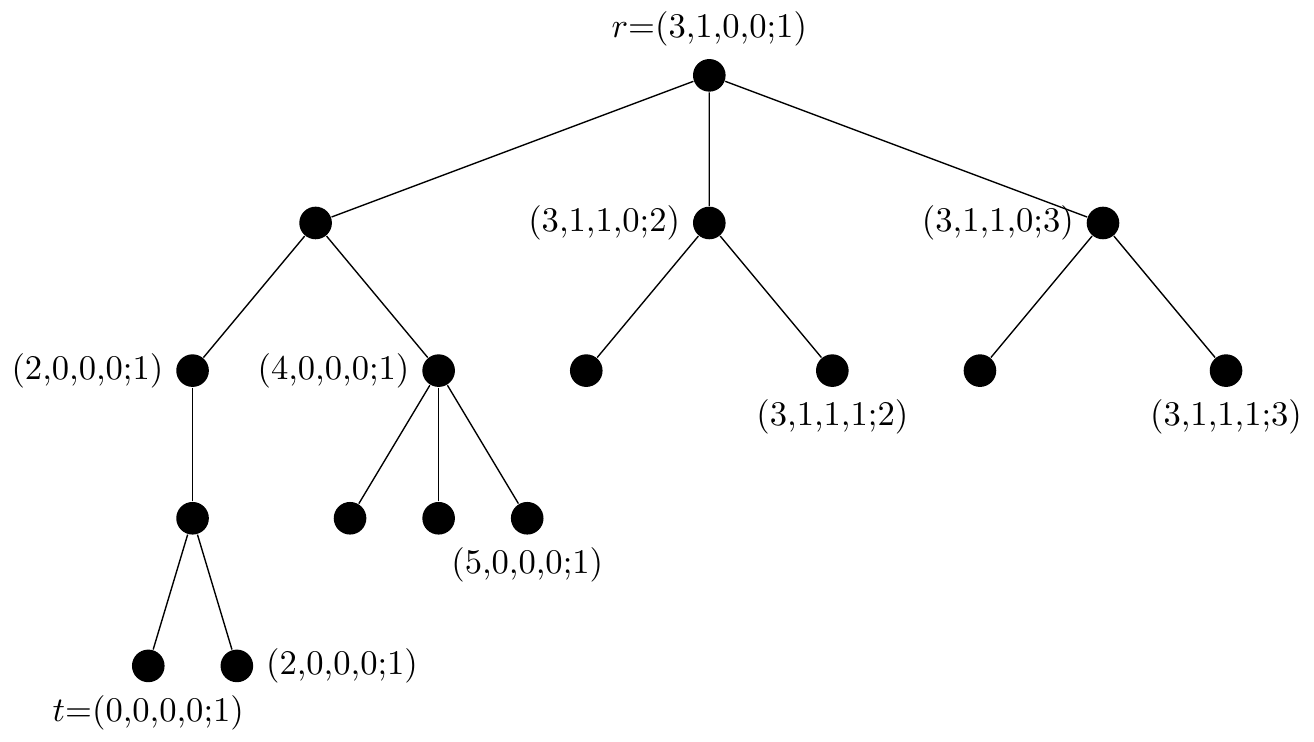}
    \caption{$5$-tuple labels for various nodes in the tree; these are obtained by finding the shortest path form $\targetnode$ and forming the path label as in fig.~\ref{fig:path_labels}. Note that the first four coordinates always sum to the edge-distance to $\targetnode$.\label{fig:node_labels}}
\end{figure}

We can now use these labels to refer to classes of nodes as well. Given a class of paths with label $(k,l,m,n;\mu)$, consider only those paths $P$ starting at $\targetnode$. We may then 
label nodes $v$ via the labels of the shortest path starting in $t$ and ending in $v$. 
Examples are shown in fig.~\ref{fig:node_labels}.

We stress that by $(k,l,m,n;\mu)$ we always refer to shortest paths, which are unique in trees, the label for each node is always well-defined in this way. However, just as paths are not uniquely identified by their label, 
the same is true for nodes labelled in the way just introduced
-- for instance, fig.~\ref{fig:node_labels} shows two nodes that can be described by $(2,0,0,0;1)$. The only nodes fully identifiable by their labels are $\target\sim(0,0,0,0;1)$ and $\rootnode\sim(h-1,1,0,0;1)$.  However, the ``resolution'' provided by these labels is sufficient for a statistical description of the pattern-distances to $\targetnode$ for each node in the graph.
This is the subject of the next appendix.

\section{Conditional Pattern-Distance along Shortest Paths}\label{app:conditional_pattern_distance}

In this appendix we derive the conditional PMF $\prob{ d^v=x \mid d^u }$
for the pattern-distance of node $v$ to $t$
given the pattern-distance to $t$ of another node $u$, located
on the shortest path between $\targetnode$ and $v$. 
To this purpose, we first derive an expression for the 
probability of the event
$\{ \xi_j^v\neq \xi_j^{\targetnode} \}$ given $d^u$, 
which can be calculated appealing to Bayes' rule.
Bayes' rule states that the conditional probability of the event $A$ given an event $B$ with $\prob{B}\neq 0$ obeys
\begin{equation}\label{eq:Bayes}
    \prob{A \mid B} = \prob{B \mid A}\frac{\prob{A}}{\prob{B}}\ ;
\end{equation}
consequently, we can write the probability of $\{ \xi_j^v\neq \xi_j^{\targetnode} \}$ given $d^u$ as
\begin{equation}\label{eq:bits_neq_conditional}
    \prob{ \xi_j^v\neq \xi_j^{\targetnode} \mid d^u=y }=\prob{  d^u=y \mid \xi_j^v\neq \xi_j^{\targetnode} }\frac{ \prob{\xi_j^v\neq \xi_j^{\targetnode}} }{ \prob{ d^u=y } }\ ,
\end{equation}
and we proceed calculating the terms on the right hand side individually.

Clearly, the pattern distance $d^u$ to the target is a sum of binomial random variables, whose statistics depends on $\Delta$. If $\Delta=\Deltamax$, all bits of the pattern $\bxi^u$ have the same expectation $\anglemean{\xi^u_j}=\ah$, and are therefore identically distributed. As the bits are independent, $d^u$ is the binomial random variable
\begin{equation}\label{eq:dv_binomial}
    d^u \sim \binomrv{f_{\Deltamax}^{\targetnode u}}{L} 
\end{equation}
with ``success probability''
\begin{equation}\label{eq:f_max_def}
    f_{\Deltamax}^{uv}:=\prob{ \xi_j^u\neq \xi_j^v  }\ .
\end{equation}
In contrast to this, $\Delta < \Deltamax$ implies that $a^u_j=\anglemean{\xi^u_j}$, and hence $\prob{ \xi_j^t\neq \xi_j^{u} }$, depends on $j$ and on the Part containing $u$ as prescribed by eq.~\eqref{eq:ah_al_def}. Therefore, instead of eq.~\eqref{eq:f_max_def} we consider
\begin{equation}\label{eq:g_def}
    g^{uv}\left( a^u_j,a^v_j \right) = \prob{\xi^v_{j}\neq \xi^{u}_j} \ .
\end{equation}
In order to extend eq.~\eqref{eq:dv_binomial} to general $\Delta$, we make the simplifying assumption that the 
bits of a given pattern are identically distributed, with probabilities averaged over all bits of that pattern. That is, we make the approximation
\begin{align}\label{eq:dv_binomial_approx}
    \nonumber
    d^u &\sim \binomrv{  f_\Delta^{\targetnode u} }{ L }\ , \\ 
    f_\Delta^{uv} &:= \sum_{x,y\in\{h,l\}} O_\Delta^{\mu\nu}(a_x, a_y) g^{uv}(a_x,a_y)\ ,
\end{align}
where $\mu$ and $\nu$ are the parts containing $u$ and $v$, respectively, and $O_\Delta^{\mu\nu}(a_x, a_y)$ is the fraction of indices $j$ such that $a^\mu_j=a_x$ while $a^\nu_j=a_y$.
By definition of $\Delta$ in sec.~\ref{sec:model}, and with the help of fig.~\ref{fig:overlap_schema}, we can write these fractions as
\begin{align}
    \nonumber
    O_\Delta^{\mu \nu}(a_h,a_h)&=
    \begin{cases}
        \frac{1}{L}\left[ \lc+2\Delta  \right] \quad &\colon \mu=\nu\ , \\ 
        \frac{1}{L}\left[ \max\left(0, (2-\nu)\lc+2\Delta \right) + \max\left( 0,\nu\lc-L+2\Delta \right)  \right] \quad &\colon \mu=1\neq\nu\ ,
    \end{cases}
    \\ \nonumber
    O_\Delta^{\mu \nu}(a_l,a_l)&=
    \begin{cases}
        \frac{1}{L}\left[ L-\lc-2\Delta \right] \quad &\colon \mu=\nu\ , \\
        \frac{1}{L}\left[ \max\left(0, (\nu-2)\lc-2\Delta \right) + \max\left( 0,L-\nu\lc-2\Delta \right) \right] \quad &\colon \mu=1\neq \nu\ ,
    \end{cases}\\ 
    O_\Delta^{\mu \nu}(a_h,a_l)&=O_\Delta^{\mu}(a_l,a_h)=\frac{1}{2}\left(1-O_\Delta^{\mu}(a_h,a_h)-O_\Delta^{\mu}(a_l,a_l)\right)\ . 
\end{align}
Notice that the definition of the $f_\Delta$'s is consistent with eq.~\eqref{eq:f_max_def} because $O_{\Deltamax}^{\mu \nu}(a_h,a_h)=1$. 
Under the assumptions of eq.~\eqref{eq:dv_binomial_approx}, the fraction in eq.~\eqref{eq:bits_neq_conditional} can be written as 
\begin{equation}\label{eq:bits_neq_conditional_fraction}
    \frac{ \prob{\xi_j^v\neq \xi_j^{\targetnode}} }{ \prob{ d^u=y } }=\frac{ f_\Delta^{\targetnode v} }{\binom{L}{y} \left(f_\Delta^{\targetnode u}\right)^{y} \left(1-f_\Delta^{\targetnode u}\right)^{L-y} }\ ,
\end{equation}
using the expression for the binomial PMF in eq.~\eqref{eq:binomial_pmf}.

To calculate the conditional probability $\prob{  d^u=y \mid \xi_j^v\neq \xi_j^{\targetnode} }$ in eq.~\eqref{eq:bits_neq_conditional}, we can use the fact that bits are independent, and consider the pattern-distance $d_{(j)}^u$ of $u$ that disregards bit $j$. This allows us to split the event $\{d^u=y\}$ into the cases where $\xi_j^u=\xi_j^{\targetnode}$ and $\xi_j^u\neq \xi_j^{\targetnode}$, respectively:
\begin{align}\label{eq:bits_neq_conditional_first_factor}
    \nonumber
    \prob{ d^u=y \mid \xi_j^v\neq \xi_j^{\targetnode} }=&
    \prob{ d^u_{(j)}=y,\xi_j^u=\xi_j^{\targetnode}  \mid \xi_j^v\neq \xi_j^{\targetnode} } \\ \nonumber &+  \prob{ d^u_{(j)}=y-1,\xi_j^u\neq \xi_j^{\targetnode}  \mid \xi_j^v\neq \xi_j^{\targetnode} } \\ \nonumber
    =&
    \prob{ d^u_{(j)}=y } \prob{\xi_j^u=\xi_j^{\targetnode}  \mid \xi_j^v\neq \xi_j^{\targetnode} } \\ 
    &+  \prob{ d^u_{(j)}=y-1} \prob{\xi_j^u\neq \xi_j^{\targetnode}  \mid \xi_j^v\neq \xi_j^{\targetnode} } \ .
\end{align}
In this expression, the marginal probabilities of $d^u_{(j)}$ are given by the binomial PMF eq.~\eqref{eq:binomial_pmf} with $n$ replaced by $L-1$ and $p=f_\Delta^{\targetnode u}$, whereas $\left\{\xi_j^u\neq \xi_j^{\targetnode}  \mid \xi_j^v\neq \xi_j^{\targetnode} \right\}$ is the same event as $\left\{\xi_j^u= \xi_j^v  \mid \xi_j^v\neq \xi_j^{\targetnode} \right\}$, which has probability $1-f_\Delta^{uv}$. We can now combine the eqs.~\eqref{eq:bits_neq_conditional_fraction} and \eqref{eq:bits_neq_conditional_first_factor} into eq.~\eqref{eq:bits_neq_conditional} to obtain
\begin{align}\label{eq:bits_neq_conditional_explicit}
    \nonumber
    \prob{ \xi_j^v\neq \xi_j^{\targetnode} \mid d^u=y }=& 
    \frac{(1-f_\Delta^{uv})f_\Delta^{\targetnode v} y}{f_\Delta^{\targetnode u}}\frac{\binom{L-1}{y-1}}{\binom{L}{y}} + \frac{f_\Delta^{uv}f_\Delta^{\targetnode v}(L-y)}{(1-f_\Delta^{ \targetnode u})}\frac{\binom{L-1}{y}}{\binom{L}{y}}\\ 
   =& \frac{(1-f_\Delta^{uv})f_\Delta^{\targetnode v} y}{f_\Delta^{\targetnode u} L} + \frac{f_\Delta^{uv}f_\Delta^{\targetnode v}(L-y)}{(1-f_\Delta^{\targetnode u})L}\ .
\end{align}

Due to the assumption of eq.~\eqref{eq:dv_binomial_approx},
this expression is independent of $j$, which implies that
$\left. d^v \mid d^u \right.$ is a binomial random variable as well, with ``success'' probability as in eq.~\eqref{eq:bits_neq_conditional_explicit}, and $L$ trials. Therefore, we can calculate the conditional expectation 
\begin{equation}\label{eq:dv_expectation_conditional}
    \anglemean{d^v \mid d^u } 
    = d^u \left( \frac{(1-f_\Delta^{uv})f_\Delta^{\targetnode v} }{f_\Delta^{\targetnode u} }-\frac{f_\Delta^{uv}f_\Delta^{\targetnode v}}{(1-f_\Delta^{ \targetnode u})} \right) + \frac{f_\Delta^{uv}f_\Delta^{\targetnode  v}}{(1-f_\Delta^{\targetnode u})}L\ .
\end{equation}
To finish this calculation we have to find the $g$'s defined in eq.~\eqref{eq:g_def}, which will determine the probabilities $f_\Delta^u$, $f_\Delta^v$ and $f_\Delta^{uv}$ by their definition in eq.~\eqref{eq:dv_binomial_approx}. For this, the label-notation introduced in app.~\ref{app:path_labels} will be useful.

We consider a pattern $\bxi^u$ at some node $u$ in Part $\mu$, and a path $P\sim (k,l,m,n;\nu)$ starting from $u$ and ending in $v$.
The transition probability from the $j$-th bit of pattern $\bxi^u$ to the $j$-th bit of the pattern at the end of the path $P$ is given by
\begin{equation}\label{eq:QP_def}
    \bQ_j^P := \left(\bQ_j^{\uparrow \mu}\right)^k \left( \bR_j^{\uparrow \mu } \right)^{l} \left( \bR_j^\nu\right)^{m} \left( \bQ_j^\nu\right)^n\ .
\end{equation}
The $\bQ$'s are as defined in eq.~\eqref{eq:Q_def}, with powers as in eq.~\eqref{eq:Q_power}. The matrix $\bR_j^\nu$ is defined as in eq.~\eqref{eq:R_def} depending only on the Part-index of the label $P$,
and that only if $\Delta < \Deltamax$; if $\Delta=\Deltamax$, the $\bR^\nu_j$ are equal for all $j$ and $\nu$.
Moreover, $\bRup$ is the family of transition matrices comprised by elements $\bR_j^{\uparrow \mu}( \xi_j \vert \xi_j^{\mu} )=P_j^{\mu}( \xi_j \vert \xi_j^{\mu}  )$ for $\xi_j^{\mu}, \xi_j \in \{0,1\}$,
which are given by Bayes' rule, eq.~\eqref{eq:Bayes}.
Consequently, the elements of $\bR_j^{\uparrow \mu}$ read
\begin{equation}\label{eq:Rup}
    \bR_j^{\uparrow \mu}=
    \begin{pmatrix}
        \frac{1-a}{1-a^{\mu}_j}(1-\Gamma^\prime) & 1-\frac{1-a}{1-a^{\mu}_j}(1-\Gamma^\prime) \\
        \frac{(1-a)\Gamma^\prime}{a^{\mu}_j} & 1-\frac{(1-a)\Gamma^\prime}{a^{\mu}_j}
    \end{pmatrix}\ .
\end{equation}
Similarly, $\left( \bQ_j^{\uparrow \mu} \right)^k$ is the matrix with elements $P_j^v( \xi_j^{\mu} \vert \xi_j^u  )$ for $\xi_j^u, \xi_j^{\mu}\in \{0,1\}$ . However, due to our stipulation that $a_j^u$ depends only on the Part $\mu$ in which $v$ is located (see eq.~\eqref{eq:identical_expectations}), a quick calculation reveals that $\bQ_j^{\uparrow u}=\bQ_j^{u}$.
Thus the probabilities $g^{P}(a^{\mu}_j,a^\nu_j)=\prob{\xi_j^u\neq \xi_j^v}$ are given by
\begin{equation}\label{eq:switching_rate_def}
    g^{(k,l,m,n;\nu)}(a_j^{\mu},a_j^\nu) :=g^{uv}(a_j^{\mu},a_j^\nu)=
    a^{\mu}_{j} Q^\nu_j(0 \vert 1) + (1-a^{\mu}_{j}) Q^\nu_j(1 \vert 0)\ ,
\end{equation}
where the dependence on $a_j^\nu$ is implicit in $\bQ_j^\nu$.
In the following we report the relevant values for $g$ by explicitly expanding eq.~\eqref{eq:switching_rate_def} in terms of the matrix elements given in eqs.~\eqref{eq:R_def}, \eqref{eq:Rup} and \eqref{eq:Q_power}.

There are essentially five different cases to consider for eq.~\eqref{eq:switching_rate_def}, each one corresponding to $P$ being one of the paths 
$(k,0,0,0;1)$, $(k,1,0,0;1)$, $(k,1,1,m;\nu)$, $(0,0,1,m;\nu)$ or $(0,0,0,m;\nu)$.
By direct calculation starting from eq.~\eqref{eq:switching_rate_def} we find the probabilities
\begin{align}\label{eq:switching_rates_explicit}
    \nonumber
    g^{(k,0,0,0;1)}(x,y)&=2x(1-x) \left[1- (1-\Gamma)^{k} \right] , \\  
    \nonumber
    g^{(k,1,0,0;1)}(x,y)&=a+x-2a x+2(1-a)(1-\Gamma)^k(\Gammap-x)\ ,
    \\
    \nonumber
    g^{(k,1,1,m;\nu)}(x,y)&=\frac{2}{a}(1-\Gamma)^{k+m}\left[ (1-a)\Gammap\left(x+y -\Gammap \right) - x y \right]+x+y-x y(1-(1-\Gamma)^{k+m}) \ ,
    \\ \nonumber
    g^{(0,0,1,m;\nu)}(x,y)&=2a(1-a)+2(1-a)(1-\Gamma)^m\left[ \Gammap-a \right] \ , \\
    g^{(0,0,0,m;\nu)}(x,y)&=2y(1-y) \left[1- (1-\Gamma)^{k} \right]\ .
\end{align}

The results of this appendix are the probabilistic building blocks required to calculate the elements of the expected transition matrix $\anglemean{\bW}$, needed in sec.~\ref{sec:compfunc}.  This calculation is shown in the next appendix~\ref{app:local_bias_mean_field}.

\section{Local Weights in the Mean Transition Matrix\label{app:local_bias_mean_field}}
This appendix combines the results of the previous appendices~\ref{app:path_labels} and \ref{app:conditional_pattern_distance} to find approximate expressions for the elements of $\anglemean{\bW}$, which are needed to derive the main result of sec.~\ref{sec:compfunc}, eq.~\eqref{eq:complexity_full}.

In what follows, we will use again the convention to enumerate the Parts of the tree in such a way that $\targetnode$ is a node descending from Part ${1}$, which we call the \emph{target-Part}. Also, given a node $v$, we will say that the neighbour closest (in terms of edge-distance) to $\targetnode$ lies in the \emph{target-wards} neighbourhood of $v$, and the neighbour closest to $\rootnode$ lies in the \emph{root-wards} neighbourhood of $v$. The remaining -- non-target -- Parts can be enumerated in any order.

\begin{figure}[h]
    \centering
    \includegraphics{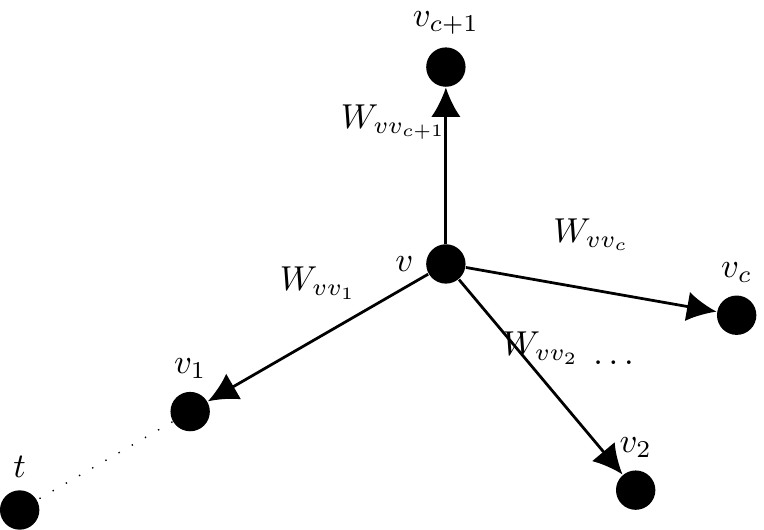}
    \caption{Neighbourhood of the node $v$, separated into target-ward, root-ward and all other neighbours. The $W$'s denote weights of edges pointing away from $v$. If $v$ is the root, then there is no root-ward neighbour $v_{c+1}$, and if $v$ is a leaf, there is \emph{only} the neighbour $v_{c+1}$.}
    \label{fig:local_bias}
\end{figure}
The derivation in this appendix is based on the observation that every node except $\targetnode$ has exactly one target-wards neighbour, enumerated as $v_1$, with corresponding edge weight $W_{v v_1}$. 
We will
denote the other neighbours of $v$ by $v_i$ with corresponding edge weights
$W_{v v_i}$ for $i>1$,
as depicted in fig.~\ref{fig:local_bias}. By convention, if the root-wards node is different from $v_1$, then $v_{c+1}$ denotes that root-wards neighbour.
Our model definitions suggest that $W_{v v_1}$ should on average exceed the other edge weights associated to $v$. However, evaluating $\anglemean{W_{v v_i}}$ is a complicated task due to the fact that it involves the pattern-distances of all nodes of the neighbourhood to normalise the $W_{v v_i}$'s (see eqs.~\eqref{eq:edge_weight} and \eqref{eq:edge_weight_function}). It is thus more convenient to calculate the ratios of such $W_i$'s,
\begin{align}\label{eq:local_bias}
    \anglemean { \frac{W_{v v_1}}{W_{v v_i}} } =: \varepsilon_i\ .
\end{align}
This can be done by temporarily assuming that the pattern-distance $d^{v_1} = d(\bxi^{v_1},\bxi^{\targetnode})$ of the target-wards neighbour is given. Then we can first calculate the conditional expectations 
\begin{equation}\label{eq:local_bias_conditional}
    \anglemean{ \frac{W_{v v_1}}{W_{v v_i}} \mid d^{v_1} }
    =
    \anglemean{  \frac{d^{v_i}+1}{d^{v_1}+1} \mid d^{v_1} }=\anglemean{  d^{v_i}+1 \mid d^{v_1} }\frac{1}{d^{v_1}+1}\ .
\end{equation}
Moreover, under the assumption of eq.~\eqref{eq:dv_binomial_approx}, 
$ d^{v_1} $ is a binomial random variable with parameters $p=f^{\targetnode v_1}$ and $n=L$ in the notation of eq.~\eqref{eq:binomial_pmf}.
Hence, the expectation of $\left[d^{v_1}+1\right]^{-1}$ is known to be \cite{Chao1972}
\begin{equation}\label{eq:binomial_negative_moment}
    \anglemean{ \frac{1}{d^{v_1}+1} }=\frac{1-\left(1-f_\Delta^{\targetnode v_1}\right)^{L+1}}{(L+1)f_\Delta^{\targetnode v_1}}\ .
\end{equation}
Therefore, we can write $\anglemean{W_{v v_1}/W_{v v_i}}$ explicitly by substituting  eqs.~\eqref{eq:local_bias_conditional} and \eqref{eq:binomial_negative_moment} into eq.~\eqref{eq:local_bias}%
\begin{align}\nonumber
    \anglemean{ \frac{W_{v v_1}}{W_{v v_i}} }
    =& \frac{(1-f_\Delta^{v_1 v_i})f_\Delta^{\targetnode v_i} }{f_\Delta^{\targetnode v_1} }-\frac{f_\Delta^{v_1 v_i}f_\Delta^{\targetnode v_i}}{(1-f_\Delta^{\targetnode v_1})} \\  
    &+ \frac{1-\left(1-f_\Delta^{\targetnode v_1}\right)^{L+1}}{(L+1)f_\Delta^{\targetnode v_1}}\left(1+\frac{f_\Delta^{v_1 v_i}f_\Delta^{\targetnode v_i}}{(1-f_\Delta^{\targetnode v_1})}L -  \frac{(1-f_\Delta^{v_1 v_i})f_\Delta^{\targetnode v_i} }{f_\Delta^{\targetnode v_1} }+\frac{f_\Delta^{v_1 v_i }f_\Delta^{\targetnode v_i}}{(1-f_\Delta^{\targetnode v_1})} \right)\ ,
\end{align}
where $f_\Delta^{\targetnode v_1}$ and $f_\Delta^{\targetnode v_i}$ are given by the probabilities described in eq.~\eqref{eq:switching_rates_explicit}, in terms of the paths connecting $v_1$ and $v_i$ to $\targetnode$, respectively. $f_\Delta^{v_1 v_i}$ is given by $f_\Delta^{P}$ with $P$ the labelled path connecting $v_1$ to $v_i$ over $v$. In the notation of eq.~\eqref{eq:local_bias}, we thus have
\begin{equation}
    \anglemean{\frac{W_{v v_1}}{W_{v v_i}}}=\varepsilon_{i}=\varepsilon\left(f_\Delta^{\targetnode v_1},f_\Delta^{\targetnode v_i}\right)
\end{equation}
with
\begin{equation}\label{eq:epsilon_def}
    \varepsilon(x,y):=\left(1+\frac{Lyz}{1-x}-\frac{y(1-z(x,y))}{1-x}+\frac{yz(x,y)}{1-x}\right)\frac{1-(1-x)^{L+1}}{(L+1)x}+\frac{y(1-z(x,y))}{x}-\frac{yz(x,y)}{1-x}\ 
\end{equation}
and $z(x,y)$ representing the $f_\Delta$ for the path connecting $v_1$ and $v_i$
\begin{equation}
    z\left( f_\Delta^{\targetnode v_1}, f_\Delta^{\targetnode v_i} \right)=f_\Delta^{v_1 v_i}\ .
\end{equation}

Having derived an expression for the ratios $\varepsilon_i=\langle W_{v v_1}/W_{v v_i}\rangle$, we now use these to compute the averages $\langle W_{v v_i}\rangle$ that we were originally interested in, by 
venturing the approximation
\begin{equation}\label{eq:bias_expectation_approximation}
    \anglemean{W_{v v_1}}=\anglemean{\frac{W_{v v_1}}{W_{v v_i}}W_{v v_i} } \approx \anglemean{W_{v v_i}}\varepsilon_{i}
\end{equation}
for all $i>1$ in the neighbourhood of $v$ (see fig.~\ref{fig:local_bias}).
Now all expected weights $\anglemean{W_{v v_i}}$ in the neighbourhood of $v$ are approximately determined by the system of equations
\begin{align}\label{eq:local_bias_lse}
    \nonumber
    \anglemean{W_{v v_1}} -\varepsilon_{i} \anglemean{W_{v v_i}} &= 0 \quad (\forall i>1)\ , \\ 
    \sum_{i\geq 1} \anglemean{ W_{v v_i} } &= 1\ .
\end{align}
The unique solution of this linear system is given by 
\begin{align}\label{eq:local_bias_gen_solution}
    \nonumber
    \anglemean{W_{v v_1}}&=\frac{1}{Z}\prod_{i>1}\varepsilon_{i}\ , \\
    \anglemean{W_{v v_i}}&=\frac{1}{Z}\prod_{j>1; j\neq i} \varepsilon_{j} \quad (i>1)\ ,
\end{align}
with $Z$
the normalising constant. 

Important specialisations of this formula are those for which (i) all $f^{\targetnode v_i}$'s and $f^{v_i v_1}$'s for $i>1$ are equal, i.e. when all $v_i$ ($i>1$) have the same path label $P_i$ relative to the target, and (ii) all $v_i$'s for ($1<i<c+1$) are labelled by the same $P_i$, but $v_{c+1}=\rootnode$, which has its own unique label. 
Case (i) applies unless $v$ is the root or the Part level node $\mu=1$. Case (ii) applies if $v$ is the Part node $\mu=1$.

In the first case, all $\anglemean{W_{v v_i}}$'s and $\varepsilon_{i}$'s for $i>1$ have to be equal, which produces
\begin{align}\label{eq:mean_weights_c_identical_edges}
    \nonumber
    \anglemean{W_{v v_1}}&\approx \frac{\varepsilon_{i}}{c+\varepsilon_{i}}\ , \\ 
    \anglemean{W_{v v_i}}&\approx \frac{1}{c+\varepsilon_{i}} \ .
\end{align}
In the second case, we have to distinguish $\anglemean{W_{v v_i}}$ for $1<i<c+1$ and $\anglemean{W_{v v_{c+1}}}$ with the result
\begin{align}\label{eq:mean_weight_c-1_identical_edges}
    \nonumber
    \anglemean{W_{v v_1}}&\approx \frac{\varepsilon_{i}\varepsilon_{c+1}}{(c-1)\varepsilon_{c+1}+\varepsilon_{i}+\varepsilon_{i}\varepsilon_{c+1}} \ ,\\ 
    \nonumber
    \anglemean{W_{v v_i}}&\approx \frac{\varepsilon_{c+1}}{(c-1)\varepsilon_{c+1}+\varepsilon_{i}+\varepsilon_{i}\varepsilon_{c+1}}\ , \\ 
    \anglemean{W_{v v_{c+1}}}&\approx \frac{\varepsilon_{i}}{(c-1)\varepsilon_{c+1}+\varepsilon_{i}+\varepsilon_{i}\varepsilon_{c+1}} \ .
\end{align}

\section{MFPT from Mean Transition Matrix\label{app:MFPT_mean_field}}

This appendix combines the findings of apps.~\ref{app:path_labels}, \ref{app:conditional_pattern_distance} and \ref{app:local_bias_mean_field} with eq.~\eqref{eq:mfpt_necklace} to state the main result of sec.~\ref{sec:compfunc} in eq.~\eqref{eq:complexity_full}.

As laid out in the previous appendices, we can approximately calculate the elements of the transition matrix, averaged over realisations of patterns. The symmetries of this approximate matrix allow us to employ a relatively simple combinatorial argument for eq.~\eqref{eq:mfpt_necklace}, where the $v_I$'s are the nodes of the path $(h-1,1,0,0;1)$ from $\target$ to $\rootnode$, though enumerated in reverse order, 
$\rootnode=v_0$, \dots, $\target=v_h$. In fact, we can express the fractions $\Pi_J/\pi_{v_J}$ in terms of sums of fractions of $\varepsilon$'s as described in the following paragraphs.

Recall the symbol $\frakt_u$ from sec.~\ref{sec:mfpt_rev} as the unique subset of directed edges pointing towards $u$. Further, with the notation introduced in sec.~\ref{sec:mfpt_rev} for dividing the tree into clusters (cf. fig.~\ref{fig:tree_clustering}), let us label the nodes within a given cluster $J$ as $v_{Ji}$, with $0\leq i \leq |J|$ and $|J|$ being the size of cluster $J$. By convention, the index $i=0$ is reserved for the node of $J$ connecting $J$ to the 
path $(h-1,1,0,0;1)$, i.e. $v_{J0}=v_J$.
For any node $v_{Ji}\in J$ the set $\frakt_{v_{Ji}}$ differs from $\frakt_{v_J}$ only by the direction of the edges between $v_J$ and $v_{Ji}$. For instance, let $J>1$; if $v_{Ji}$ is an immediate descendant of $v_J$, then the edge-distance between $\target$ and $v_J$ is $h-J$, and
\begin{align}
    \frac{\pi_{v_{Ji}}}{\pi_{v_J}}=
    \frac{c+\varepsilon\left( f_\Delta^{(h-J,0,0,0;1)},f_\Delta^{(h-J+2,0,0,0;1)} \right)}{\varepsilon\left( f_\Delta^{(h-J,0,0,0;1)},f_\Delta^{(h-J+2,0,0,0;1)} \right)\left[c+\varepsilon\left( f_\Delta^{(h-J-1,0,0,0;1)},f_\Delta^{(h-J+1,0,0,0;1)} \right) \right]  }\ ,
\end{align}
using eq.~\eqref{eq:mean_weights_c_identical_edges} and $\varepsilon$ defined as in eq.~\eqref{eq:epsilon_def}. We also utilised that in this instance, the shortest paths from $\target$ to $v_J$ and $v_{Ji}$ have the form $(h-J,0,0,0;1)$ and $(h-J+1,0,0,0;1)$, respectively.

In $\Pi_J/\pi_{v_J}$ this term appears as a summand $c-1$ times, because $v_J$ has that many immediate descendants that are not on the path to the target, i.e. that have label $(h-J+1,0,0,0;1)$. Repeating this analysis for all $h-J$ lower levels of the cluster $J$ (where there are now $c$ immediate descendants to each node that is not a leaf), we find the expression
\begin{align}\label{eq:PiJ/piJ}
\nonumber
    \frac{\Pi_J}{\pi_{v_J}} =& 1+ \frac{c-1}{c+\varepsilon\left( f_\Delta^{(h-J-1,0,0,0;1)},f_\Delta^{(h-J+1,0,0,0;1)} \right)} \\ \nonumber
    \times&\left[
    \sum_{\ell=1}^{h-J-1} c^{\ell-1} \frac{ c+\varepsilon\left( f_\Delta^{(h-J-1+\ell,0,0,0;1)},f_\Delta^{(h-J+1+\ell,0,0,0;1)} \right) }{ \prod_{k=1}^{\ell}\varepsilon\left( f_\Delta^{(h-J-1+k,0,0,0;1)},f_\Delta^{(h-J+1+k,0,0,0;1)} \right) } \right. \\ 
    &\left. + \frac{ c^{h-J-1} }{ \prod_{k=1}^{h-J-1}\varepsilon\left( f_\Delta^{(h-J-1+k,0,0,0;1)},f_\Delta^{(h-J+1+k,0,0,0;1)} \right) }\right].
\end{align}
The last summand within the brackets arises from the fact that all leaves have only one outgoing edge. 

If $J=1$, then the mean edge weights at $v_J$ are given by eq.~\eqref{eq:mean_weight_c-1_identical_edges}, whereas lower edges inside the cluster still follow eq.~\eqref{eq:mean_weights_c_identical_edges},
\begin{align}\label{eq:Pi1/pi1}
    \nonumber
    &\frac{\Pi_{1}}{\pi_{v_{1}}} = 1+\frac{(c-1)\varepsilon\left( f_\Delta^{(h-2,0,0,0;1)},f_\Delta^{(h-1,1,0,0;1)} \right)}{
    d_1} \\ 
    &\times \left[ \sum_{\ell=1}^{h-2} c^{\ell-1} \frac{ c+\varepsilon\left( f_\Delta^{(h-2+\ell,0,0,0;1)},f_\Delta^{(h+\ell,0,0,0;1)} \right) }{ \prod_{k=1}^{\ell}  \varepsilon\left( f_\Delta^{(h-2+k,0,0,0;1)},f_\Delta^{(h+k,0,0,0;1)} \right) } +  \frac{ c^{h-2} }{ \prod_{k=1}^{h-2}  \varepsilon\left( f_\Delta^{(h-2+k,0,0,0;1)},f_\Delta^{(h+k,0,0,0;1)} \right) }\right],
\end{align}
with the denominator
\begin{align}
    \nonumber
    d_1=&(c-1)\varepsilon\left( f_\Delta^{(h-2,0,0,0;1)},f_\Delta^{(h-1,1,0,0;1)} \right)+\varepsilon\left( f_\Delta^{(h-2,0,0,0;1)},f_\Delta^{(h,0,0,0;1)} \right)\\ 
    &+\varepsilon\left( f_\Delta^{(h-2,0,0,0;1)},f_\Delta^{(h-1,1,0,0;1)} \right)\varepsilon\left( f_\Delta^{(h-2,0,0,0;1)},f_\Delta^{(h,0,0,0;1)} \right)
\end{align}
Finally, for $J=0$ we have $v_0=\rootnode$, which brings us back to eq.~\eqref{eq:local_bias_gen_solution} for the edges connecting to $v_j=\rootnode$. 
This observation leads us to
\begin{align}\label{eq:Pi0/pi0}
    \nonumber
     \frac{\Pi_0}{\pi_{v_0}} =& 1+ \frac{1 }{\prod_{\mu=2}^{c}\varepsilon\left( f_\Delta^{(h-1,0,0,0;1)},f_\Delta^{(h-1,1,1,0;\mu)} \right)+\sum_{\mu=2}^{c}\prod_{\nu=2; \nu\neq\mu}^{c}\varepsilon\left(f_\Delta^{(h-1,0,0,0)},f_\Delta^{(h-1,1,1,0;\nu)}\right) } \\ \nonumber
    \times&
    \sum_{\mu=2}^c\prod_{\nu=2; \nu\neq\mu}^{c}\varepsilon\left(f_\Delta^{(h-1,0,0,0)},f_\Delta^{(h-1,1,1,0;\nu)}\right)
    \left[
    \frac{ c+\varepsilon\left( f_\Delta^{(h-1,1,0,0;1)},f_\Delta^{(h-1,1,1,1;\mu)} \right) }{ \varepsilon\left( f_\Delta^{(h-1,1,0,0;1)},f^{(h-1,1,1,1;\mu)} \right)} \right. \\ \nonumber
    &\qquad +\left.
    \frac{ c+\varepsilon\left( f_\Delta^{(h-1,1,1,0;\mu)},f_\Delta^{(h-1,1,1,2;\mu)} \right) }{ \varepsilon\left( f_\Delta^{(h-1,1,0,0;1)},f_\Delta^{(h-1,1,1,1;\mu)} \right)\varepsilon\left( f_\Delta^{(h-1,1,1,0;\mu)},f_\Delta^{(h-1,1,1,2;\mu)} \right)}
    \right. \\ 
    &\qquad +\left.
    \sum_{\ell=3}^{h-1} c^{\ell-1} \frac{ c+\varepsilon\left( f_\Delta^{(h-1,1,1,l-2;\mu)},f_\Delta^{(h-1,1,1,l;\mu)} \right) }{
    d^\mu_{0,\ell} }
    + 
    \frac{ c^{h-1} }{
    d^\mu_{0,h-1}    }\right]\ 
\end{align}
with the denominator terms
\begin{align}
    \nonumber
    d^\mu_{0,\ell}=&
    \varepsilon\left( f_\Delta^{(h-1,1,0,0;1)},f_\Delta^{(h-1,1,1,1;\mu)} \right)\varepsilon\left( f_\Delta^{(h-1,1,1,0;\mu)},f_\Delta^{(h-1,1,1,2;\mu)} \right)\\ 
    &\times\prod_{k=3}^{\ell}\varepsilon\left( f_\Delta^{(h-1,1,1,k-2;\mu)},f_\Delta^{(h-1,1,1,k;\mu)} \right) 
\end{align}

For eq.~\eqref{eq:mfpt_necklace}, we need to combine these expression with appropriate path weights connecting the clusters. More precisely, we need the fractions
\begin{equation}
    \frac{ \anglemean{W_{v_{I-1}v_{I-2}}} \dots \anglemean{W_{v_{K+1}v_K}} } { \anglemean{W_{v_K,v_{K+1}}} \dots \anglemean{W_{v_{I-1}v_{I}}} }=\frac{ \anglemean{W_{v_{I-1}v_{I-2}}} }{ \anglemean{W_{v_{I-1}v_I}} }\frac{ \anglemean{W_{v_{I-2}v_{I-3}}} }{ \anglemean{W_{v_{I-2}v_{I-1}}} }\cdots \frac{ \anglemean{W_{v_{K+1}v_K}} }{ \anglemean{W_{v_{K+1}v_{K+2}}} }\frac{1}{ \anglemean{W_{v_K v_{K+1}}} }\ .
\end{equation}
If $K>1$ this fraction can be written as
\begin{align}
    \nonumber
    \frac{ \anglemean{W_{v_{I-1}v_{I-2}}} \dots \anglemean{W_{v_{K+1}v_K}} } { \anglemean{W_{v_K,v_{K+1}}} \dots \anglemean{W_{v_{I-1}v_{I}}} }=&\left[c+\varepsilon\left( f_\Delta^{(h-K-1,0,0,0;1)},f_\Delta^{(h-K+1,0,0,0;1)} \right) \right]\\ 
    &\times \prod_{J=K}^{I-1}\frac{1}{\varepsilon\left( f_\Delta^{(h-J-1,0,0,0;1)},f_\Delta^{(h-J+1,0,0,0;1)} \right)}\ .
\end{align}
Due to the distinct form of the weights close to $\rootnode$, the same fractions for $K=1$ and $K=0$ take the form
\begin{align}
    \nonumber
    \frac{ \anglemean{W_{v_{I-1}v_{I-2}}} \dots \anglemean{W_{v_{2}v_1}} }{ \anglemean{W_{v_1,v_{2}}}\dots \anglemean{W_{v_{I-1}v_{I}}} }=
    &\frac{1}{\varepsilon\left( f_\Delta^{(h-2,0,0,0;1)},f_\Delta^{(h-1,1,0,0;1)} \right)\varepsilon\left( f_\Delta^{(h-2,0,0,0;1)},f_\Delta^{(h,0,0,0;1)} \right)}\\ \nonumber
    &\times\left( (c-1)\varepsilon\left( f_\Delta^{(h-2,0,0,0;1)},f_\Delta^{(h-1,1,0,0;1)} \right)+\varepsilon\left( f_\Delta^{(h-2,0,0,0;1)},f_\Delta^{(h,0,0,0;1)} \right)\right. \\ \nonumber
    &\quad +\left. \varepsilon\left( f_\Delta^{(h-2,0,0,0;1)},f_\Delta^{(h-1,1,0,0;1)} \right)\varepsilon\left( f_\Delta^{(h-2,0,0,0;1)},f_\Delta^{(h,0,0,0;1)} \right)  \right)
    \\ 
    &\times \prod_{J=2}^{I-1}\frac{1}{\varepsilon\left( f_\Delta^{(h-J-1,0,0,0;1)},f_\Delta^{(h-J+1,0,0,0;1)} \right)} \\ \nonumber
\end{align}
and
\begin{align}\label{eq:path_weight_ratio_v0}
    \nonumber
    &\frac{ \anglemean{W_{v_{I-1}v_{I-2}}} \dots \anglemean{W_{v_{1}v_0}} }{ \anglemean{W_{v_0,v_{1}}}\dots \anglemean{W_{v_{I-1}v_{I}}} }=\\ \nonumber
     &\frac{\prod_{\mu=2}^c\varepsilon\left( f_\Delta^{(h-1,0,0,0;1)},f_\Delta^{(h-1,1,1,0;\mu)} \right)+\sum_{\mu=2}^c\prod_{\nu=2;\nu\neq\mu}^{c}\varepsilon\left( f_\Delta^{(h-1,0,0,0;1)},f_\Delta^{(h-1,1,1,0;\nu)} \right)  }{\varepsilon\left( f_\Delta^{(h-2,0,0,0;1)},f_\Delta^{(h-1,1,0,0;1)}\right) \varepsilon\left( f_\Delta^{(h-2,0,0,0;1)},f_\Delta^{(h,0,0,0;1)}\right) \prod_{\mu=2}^c\varepsilon\left( f_\Delta^{(h-1,0,0,0;1)},f_\Delta^{(h-1,1,1,0;\mu)} \right)}\\ 
            &\times \prod_{J=2}^{I-1}\frac{1}{\varepsilon\left( f_\Delta^{(h-J-1,0,0,0;1)},f_\Delta^{(h-J+1,0,0,0;1)} \right)} \ ,
\end{align}
respectively.
Combining these expressions in the manner of eq.~\eqref{eq:mfpt_necklace} produces the function ${\CMF}$ shown in sec.~\ref{sec:compfunc}.

\end{document}